\documentclass[10pt,twocolumn]{article}

\usepackage{bbold}
\usepackage{amsmath,amssymb,color,enumerate}
\usepackage{graphicx}
\usepackage{calligra}
\usepackage{mathtools}
\usepackage{algorithm}
\usepackage{algcompatible}
\DeclareMathOperator*{\argmax}{arg\,max}
\DeclareMathOperator*{\argmin}{arg\,min}

\usepackage{bbm}
\usepackage[margin= 0.9in]{geometry}

\usepackage{xcolor}
\usepackage[hidelinks]{hyperref}
\newtheorem{Prop}{Proposition}

\newtheorem{Theorem}{Theorem}

\usepackage[round]{natbib}
\begin{document}

\title{Learning Mixtures of Smooth Product Distributions:\\ Identifiability and Algorithm}

\author{ \quad \quad \quad \quad Nikos Kargas \hspace{35px} Nicholas D. Sidiropoulos \\  University of Minnesota  \hspace{5px}  \quad  \quad  University of Virginia}
\date{}
\maketitle
\thispagestyle{empty}

\begin{abstract}

We study the problem of learning a mixture model of non-parametric product distributions. The problem of learning a mixture model is that of finding the component distributions along with the mixing weights using observed samples generated from the mixture. The problem is well-studied in the parametric setting, i.e., when the component distributions  are members of a parametric family -- such as Gaussian distributions. In this work, we focus on multivariate mixtures of non-parametric product distributions and propose a two-stage approach which recovers the component distributions of the mixture under a smoothness condition. 
Our approach builds upon the identifiability properties of the canonical polyadic (low-rank) decomposition of tensors, in tandem with Fourier and Shannon-Nyquist sampling staples from signal processing. We demonstrate the effectiveness of the approach on synthetic and real datasets.
\end{abstract}

\section{Introduction}

Learning mixture models is a fundamental problem in statistics and machine learning having numerous applications such as density estimation and clustering. In this work, we consider the special case of mixture models whose component distributions factor into the product of the associated marginals. An example is a mixture of axis-aligned Gaussian distributions, an important class of Gaussian Mixture Models (GMMs). Consider a scenario where different diagnostic tests are applied to patients, and test results are assumed to be independent conditioned on the binary disease status of the patient which is the latent variable. The joint Probability Density Function (PDF) of the tests can be expressed as a weighted sum of two components, and each component factors into the product of univariate marginals.  Fitting a mixture model to an unlabeled dataset allows us to cluster the patients into two groups by determining the value of the latent variable using the Maximum a Posteriori (MAP) principle.

Most of the existing literature in this area has focused on the fully-parametric setting,  where the mixture components are members of a parametric family, such as Gaussian distributions. The most popular algorithm for learning a parametric mixture model is Expectation Maximization (EM)~\citep{Dempster1977}. Recently, methods based on tensor decomposition and particularly the Canonical Polyadic Decomposition (CPD) have  gained popularity as an alternative to EM for learning various latent variable models~\citep{AnGeHsu2014b}. What makes the CPD a powerful tool for data analysis is its identifiability properties, as the CPD of a tensor is unique under relatively mild rank conditions~\citep{SiDeFu2017}.

In this work we propose a two-stage approach based on tensor decomposition for recovering the conditional densities of mixtures of smooth product distributions. We show that when the unknown conditional densities are approximately band-limited it is possible to uniquely identify and recover them from partially observed data. The key idea is to jointly factorize histogram estimates of lower-dimensional PDFs that can be easily and reliably estimated from observed samples. The conditional densities can then be recovered using an interpolation procedure. We formulate the problem as a coupled tensor factorization and propose an alternating-optimization algorithm. We demonstrate the effectiveness of the approach on both synthetic and real data.

\textbf{Notation}: Bold, lowercase, $\mathbf{x}$, and uppercase letters, $\mathbf{X}$, denote vectors and matrices respectively. Bold, underlined, uppercase letters, $\underline{\mathbf{X}}$, denote $N$-way (${N \geq 3}$) tensors. We use the notation $\mathbf{x}[i]$, $\mathbf{X}[i,j]$, $\underline{\mathbf{X}}[i,j,k]$ to refer to specific elements of a vector, matrix and tensor respectively. 
We denote the vector obtained by vertically stacking the columns of the tensor $\underline{\mathbf{X}}$ into a vector by $\text{vec}(\underline{\mathbf{X}})$. Additionally, $\textrm{diag}(\mathbf{x}) \in \mathbb{R}^{M \times M}$ denotes the diagonal matrix with the elements of vector $\mathbf{x}  \in \mathbb{R}^{M}$ on its diagonal. The set of integers $\{1,\ldots,N\}$ is denoted as $[N]$. Uppercase, $X$, and lowercase letters, $x$, denote scalar random variables and realizations thereof, respectively. 

\section{Background}

\subsection{Canonical Polyadic Decomposition}
In this section, we briefly introduce basic concepts related to tensor decomposition. An $N$-way tensor ${\underline{\mathbf{X}} \in \mathbb{R}^{I_1 \times I_2 \times \cdots \times I_N}}$ 
is a multidimensional array whose elements are indexed by $N$ indices. A polyadic decomposition  expresses $\underline{\mathbf{X}}$  as a sum of rank-$1$ terms
\begin{equation}
\underline{\mathbf{X}} = \sum_{r=1}^R\mathbf{A}_1[:,r] \circ \mathbf{A}_2[:,r] \circ \cdots  \circ \mathbf{A}_N[:,r],
\label{eq:cpd}
\end{equation}
where $\mathbf{A}_n \in \mathbb{R}^{I_n \times R}$, $ 1 \leq r\leq R$, $\mathbf{A}_n[:,r]$ denotes the $r$-th column of matrix $\mathbf{A}_n$ and $\circ$ denotes the outer product. 
If the number of rank-$1$ terms is minimal, then Equation~\eqref{eq:cpd} is called the CPD of $\underline{\mathbf{X}}$ and $R$ is called the rank of $\underline{\mathbf{X}}$. 
Without loss of generality, we can restrict the columns of $\{\mathbf{A}_n\}_{n=1}^N$ to have unit norm and have the following equivalent expression
\begin{equation}
\underline{\mathbf{X}} = \sum_{r=1}^R{ \boldsymbol{\lambda}}[r]\mathbf{A}_1[:,r]\circ \mathbf{A}_2[:,r] \circ \cdots  \circ \mathbf{A}_N[:,r],
\label{eq:CPD}
\end{equation}
where $\|{\bf A}_n[:,r]\|_p=1$ for {a certain} $p\geq 1$,  $\forall \;  n,r$, and ${ \boldsymbol{\lambda}}= \left [{ \boldsymbol{\lambda}}[1],\ldots,{ \boldsymbol{\lambda}}[R] \right]^T$ `absorbs' the norms of columns. For
convenience, we use the shorthand notation ${\underline{\mathbf{X}} = [\![ { \boldsymbol{\lambda}}, \mathbf{A}_1,\ldots,\mathbf{A}_N ]\!]}_R$.
We can express the CPD of a tensor in a matricized form. With $\odot$ denoting the Khatri-Rao (columnwise Kronecker) matrix product, it can be shown that the mode-$n$ matrix unfolding of $\underline{\mathbf{X}}$ is given by
\begin{equation}
{\mathbf{X}}^{(n)} =  \left( \underset{k \neq n}{ \underset{k=1}{ \overset{N}{\odot}}} \mathbf{A}_k \right) \textrm{diag}(\boldsymbol{\lambda}) \mathbf{A}_n^T,
\end{equation}
where
$\underset{k\neq n}{ \underset{k=1}{ \overset{N}{\odot}}} \mathbf{A}_k = \mathbf{A}_N \odot \cdots \odot \mathbf{A}_{n+1} \odot \mathbf{A}_{n-1} \odot \cdots \odot \mathbf{A}_1.$
The CPD can be expressed in a vectorized form as
\begin{equation}
\text{vec}(\underline{\mathbf{X}})=  \left( { \underset{n=1}{ \overset{N}{\odot}}} \mathbf{A}_n \right) \boldsymbol{\lambda}.
\end{equation}

It is clear that the rank-$1$ terms can be arbitrarily permuted without affecting the decomposition. We say that a CPD of a tensor is unique when it is only subject to this trivial indeterminacy. 

\subsection{Learning Problem}

Let $\mathcal{X} = \{X_n\}_{n=1}^N$ denote a set of $N$ random variables. We say that a PDF $f_{\mathcal{X}}$ is a mixture of $R$ component distributions if it can be expressed as a weighted sum of $R$ multivariate distributions
\begin{equation}
f_{\mathcal{X}}(x_1,\ldots,x_N) = \sum_{r=1}^R w_r f_{\mathcal{X}|H}(x_1,\ldots,x_N |r),
\end{equation}
where $f_{\mathcal{X}|H}$ are conditional PDFs and $\{w_r\}_{r=1}^R$ are non-negative numbers such that $\sum_{r=1}^R w_r = 1$, called mixing weights. When each conditional PDF factors into the product of its marginal densities we have that
\begin{equation}
f_{\mathcal{X}}(x_1,\ldots,x_N) = \sum_{r=1}^R w_r \prod_{n=1}^N f_{X_n|H}(x_n | r),
\label{eq:mixture_of_product}
\end{equation}
which can be seen as a continuous extension of the CPD model of Equation~\eqref{eq:CPD}. A sample from the  mixture model is generated by first drawing a component $r$ according to $w$ and then independently drawing samples for every variable $\{X_n\}_{n=1}^N$ from the conditional PDFs $ f_{X_n|H}(\cdot |r)$. The problem of learning the mixture is that of finding the conditional PDFs as well as the mixing weights given observed samples.

\subsection{Related Work}
Mixture models have numerous applications in statistics and machine learning including clustering and density estimation to name a few~\citep{McLachlan2000}. A common assumption made in  multivariate mixture models is a parametric form of the conditional PDFs. For example, when the conditional PDFs are assumed to be Gaussian, the goal is to recover the mean vectors and covariance matrices defining each multivariate Gaussian component and the mixing weights. Other common choices include categorical, exponential, Laplace or Poisson distributions. The most popular algorithm for learning the parameters of the mixture is the EM algorithm~\citep{Dempster1977} which maximizes the likelihood function with respect to the parameters. EM-based methods have been also considered for  learning  mixture models of non-parametric distributions\footnote{The term non-parametric is used to describe the case in which no assumptions are made about the form of the conditional densities.} by parameterizing the unknown conditional PDFs using kernel density estimators~\citep{Benaglia2009,levine2011}, which lack however theoretical guarantees. 

Tensor decomposition methods can be used as an alternative to EM for learning  various latent variable models~\citep{AnGeHsu2014b}. High-order moments of several probabilistic models can be expressed using low-rank CPDs. Decomposing these tensors reveals the true parameters of the probabilistic models. In the absence of noise and model mismatch, algebraic algorithms can be applied to compute the CPD under certain conditions, see~\citep{SiDeFu2017} and references therein, and \citep{hsu2013} for the application to GMMs. Tensor decomposition approaches have been proposed for learning mixture models but are mostly restricted to Gaussian or categorical distributions~\citep{hsu2013,jain2014,gottesman2018}. In practice, mainly due to sampling noise the result of these algorithms may not be satisfactory and EM can be used for refinement~\citep{zhang2014,ruffini2017}. In the case of non-parametric mixtures of product distributions, identifiability of the components has been established in~\citep{allman2009}. The authors have shown that it is possible to identify the conditional PDFs given the true joint PDF, if the conditional PDFs of each $X_n$ across different mixture components are linearly independent i.e., the continuous factor ``matrices'' have linearly independent columns. However, the exact true joint PDF is never given -- only samples drawn from it are available in practice, and elements may be missing from any given sample. Furthermore, \citep{allman2009} did not provide an estimation procedure, which limits the practical appeal of an interesting theoretical contribution. 

In this work, we focus on mixtures of product distributions of continuous variables and do not specify a parametric form of the conditional density functions. We show that it is possible to recover mixtures of \textit{smooth} product distributions from observed samples.  The key idea is to first transform the problem to that of learning a mixture of categorical distributions by decomposing lower-dimensional and (possibly coarsely) discretized joint PDFs. Given that the conditional PDFs are (approximately) band-limited (smooth), they can be recovered from the discretized PDFs under certain conditions.

\section{Approach}
Our approach consists of two stages. We express the problem as a tensor factorization problem and show that if $N\geq 3$, we can recover points of the unknown conditional CDFs. Under a smoothness condition, these points can be used to recover the true conditional PDFs using an interpolation procedure. 
\subsection{Problem Formulation}
We assume that we are given $M$ $N$-dimensional samples  $ \{ \mathbf{x}_m\}_{m=1}^M$ that have been generated from a mixture of product distributions as in Equation~\eqref{eq:mixture_of_product}.
We discretize  each random variable $X_n$ by partitioning its support into $I$ uniform intervals $\{ \Delta_n^{i} = \bigl (d_n^{i-1},d_n^{i} \bigr) \}_{ 1\leq i \leq I}$. Specifically, we consider a discretization of the PDF and define the probability tensor (histogram) $\underline{\mathbf{X}}[i_1,\ldots,i_N] \triangleq {\sf Pr} \left( X_1 \in \Delta_n^{i_1} ,\ldots,X_N \in \Delta_n^{i_N}   \right)$  given by
\begin{multline}
 \underline{\mathbf{X}}[i_1,\ldots,i_N] =
\sum_{r=1}^R w_r \prod_{n=1}^N \int_{  \Delta_n^{i_n} }  f_{X_n|H}(x_n | r) dx_n \\
= \sum_{r=1}^R w_r \prod_{n=1}^N   {\sf Pr} \left( X_n \in \Delta_n^{i_n} \right |  H = r).
\label{eq:discrete}
\end{multline}
Let $\mathbf{A}_n[i_n,r] \triangleq {\sf{Pr}}\left( X_n \in \Delta_n^{i_n} \right | H = r)$, ${\boldsymbol{\lambda}[r] \triangleq w_r}$.
Note that $\underline{\mathbf{X}}$ is an $N$-way tensor and admits a CPD with non-negative factor matrices $\{\mathbf{A}_n\}_{n=1}^N$ and rank $R$, i.e., ${\underline{\mathbf{X}} = [\![ {\boldsymbol{ \lambda}}, \mathbf{A}_1,\ldots,\mathbf{A}_N ]\!]}_R$. From equation~\eqref{eq:discrete} it is clear that the discretized conditional PDFs  are identifiable and can be recovered by decomposing the true joint discretized probability tensor, if $N \geq 3$ and $R$ is small enough,  by virtue of the uniqueness properties of CPD~\citep{SiDeFu2017}. 

In practice we do not observe ${\underline{\mathbf{X}}}$ but we have to deal with perturbed versions. Based on the observed samples, we can compute an approximation of the probability tensor $\underline{\mathbf{X}}$ by counting how many samples fall into each bin and normalizing the tensor by dividing with the total number of samples.
The size of the probability tensor grows exponentially with the number of variables and therefore the estimate  will be highly inaccurate even when the number of discretization intervals is small. More importantly, datasets often contain missing data and therefore its impossible to construct such tensor. On the other hand, it may be possible to estimate low-dimensional discretized joint PDFs of subsets of the random variables which correspond to low-order tensors. For example, in the clustering example given in the introduction some patients may be tested on a subset of the available tests. Finally, the model of Equation~\eqref{eq:discrete} is just an approximation of our original model, as our ultimate goal is to recover the true conditional PDFs. To address the aforementioned challenges we have to answer the following two questions
\begin{enumerate}
\item Is it possible to learn the mixing weights and discretized conditional PDFs from missing/limited data?
\item Is it possible to recover non-parametric conditional PDFs from their discretized counterparts?
\end{enumerate}
Regarding the first question, it has been recently shown that a joint Probability Mass Function (PMF) of a set of random variables can be  recovered from lower-dimensional joint PMFs if the joint PMF has low enough rank~\citep{KaSiFu2018}. This result allows us to recover the discretized conditional PDFs from low-dimensional histograms but cannot be extended to the continuous setting in general because of the loss of information induced from the discretization step. We further discuss and provide conditions under which we can overcome these issues.

\subsection{Identifiability using Lower-dimensional Statistics}
\label{sec:ident}
In this section we provide insights regarding the first issue. It turns out that realizations of subsets of only three random variables are sufficient to recover ${\sf{Pr}}\left( X_n \in \Delta_n^{i_n} \right | H = r)$ and $\{w_r\}_{r=1}^R$. Under the mixture model~\eqref{eq:mixture_of_product}, a histogram of any subset of three random variables $X_{j},X_{k},X_{\ell}$ denoted as $\underline{\mathbf{X}}_{jk\ell}$, with $\underline{\mathbf{X}}_{jk\ell}[i_j,i_k,i_\ell] = {\sf Pr}( X_j \in \Delta_j^{i_j}, X_k \in \Delta_k^{i_k}, X_\ell \in \Delta_\ell^{i_\ell})$ can be written as $\underline{\mathbf{X}}_{jk\ell}[i_j,i_k,i_{\ell}] = \sum_{r=1}^R \boldsymbol{\lambda} [r] \mathbf{A}_j[i_j,r] \mathbf{A}_k[i_k,r] \mathbf{A}_{\ell}[i_{\ell},r],$
which is a third-order tensor of rank $R$. A  fundamental result on the uniqueness of tensor decomposition of third-order tensors was given by in~\citep{Kru1977}. The result states that if ${\underline{\mathbf{X}}}$ admits a decomposition ${\underline{\mathbf{X}} = [\![{\boldsymbol \lambda}, \mathbf{A}_1,\mathbf{A}_2,\mathbf{A}_3 ]\!]_R}$, with ${k_{\mathbf{A}_1} + k_{\mathbf{A}_2} + k_{\mathbf{A}_3} \geq 2R + 2}$ then $\textrm{rank}(\underline{\mathbf{X}}) = R $ and the decomposition of $\underline{\mathbf{X}}$ is unique. 
Here, $k_{\mathbf{A}}$ denotes the Kruskal rank of the matrix $\mathbf{A}$ which is equal to the largest integer such that every subset of $k_{\mathbf{A}}$ columns are linearly independent. 
When the rank is small and the decomposition is exact, the parameters of the CPD model can be computed exactly via Generalized Eigenvalue Decomposition (GEVD) and related algebraic~ algorithms~\citep{leurgans1993,domanov2014,SiDeFu2017}. 
\begin{Theorem}\label{thm:leurgans}
\citep{leurgans1993} 
Let ${\underline{\mathbf{X}}}$ be a tensor that admits a polyadic decomposition  ${\underline{\mathbf{X}} = [\![{\boldsymbol \lambda}, \mathbf{A}_1,\mathbf{A}_2,\mathbf{A}_3 ]\!]_R}$, $\mathbf{A}_1\in \mathbb{R}^{I_1 \times R}$, $\mathbf{A}_2\in \mathbb{R}^{I_2 \times R}$, $\mathbf{A}_3\in \mathbb{R}^{I_3 \times R}$, $\boldsymbol{\lambda} \in \mathbb{R}^R$ and suppose that $\mathbf{A}_1$, $\mathbf{A}_2$ are full column rank and  $k_{\mathbf{A}_3} \geq 2$. Then $\textrm{rank}(\underline{\mathbf{X}}) = R $, the decomposition of $\underline{\mathbf{X}}$ is unique and can be found algebraically. 
\end{Theorem}

More relaxed uniqueness conditions from the field of algebraic geometry have been proven in recent years. 
\begin{Theorem}\label{thm:generic}
\citep{ChiOtta2012} Let ${\underline{\mathbf{X}}}$ be a tensor that admits a polyadic decomposition $\underline{\mathbf{X}} = [\![{\boldsymbol \lambda}, \mathbf{A}_1,\mathbf{A}_2,\mathbf{A}_3 ]\!]$, where $\mathbf{A}_1 \in \mathbb{R}^{I_1 \times F}$, $\mathbf{A}_2 \in \mathbb{R}^{I_2 \times F}$, $\mathbf{A}_3 \in \mathbb{R}^{I_3 \times F}$, $ I_1 \leq I_2 \leq I_3$. Let $\alpha,\beta$ be the largest integers such that $2^\alpha \leq I_1$ and $2^\beta \leq I_2$. If $F \leq 2^{\alpha + \beta -2}$ then the decomposition of $\underline{\mathbf{X}}$ is essentially unique almost surely.
\end{Theorem}
Theorem~\ref{thm:generic} is a generic uniqueness result i.e, all non-identifiable parameters form a set of Lebesgue measure zero. To see how the above theorems can be applied in our setup, consider the joint decomposition of the probability tensors $\underline{\mathbf{X}}_{jk\ell}$. Let $\mathcal{S}_1$, $\mathcal{S}_2$, and $\mathcal{S}_3$ denote disjoint ordered subsets of $[N]$, with cardinality $c_1 = |\mathcal{S}_1|$, $c_2 = |\mathcal{S}_2|$, and $c_3 = |\mathcal{S}_3|$, respectively. Let $\underline{\mathbf{Y}}$ be the $c_1\times c_2 \times c_3$ block tensor whose $(j,k,\ell)$-th block is the tensor $\underline{\mathbf{X}}_{jk\ell}$, $j \in \mathcal{S}_1$, $k \in \mathcal{S}_2$, $\ell \in \mathcal{S}_{3}$. It is clear that the tensor $\underline{\mathbf{Y}}$ admits a CPD  $\underline{\mathbf{Y}} = [\![ {\boldsymbol{ \lambda}}, \widehat{\mathbf{A}}_1, \widehat{\mathbf{A}}_2,\widehat{\mathbf{A}}_3 ]\!]_R $ where $\widehat{\mathbf{A}}_1 = [\mathbf{A}_{\mathcal{S}_1(1)}^T, \cdots, \mathbf{A}_{\mathcal{S}_1(c_1)}^T ]^T $, $\widehat{\mathbf{A}}_2 = [\mathbf{A}_{\mathcal{S}_2(1)}^T, \cdots, \mathbf{A}_{\mathcal{S}_2(c_2)}^T ]^T $, $\widehat{\mathbf{A}}_3 = [\mathbf{A}_{\mathcal{S}_3(1)}^T, \cdots, \mathbf{A}_{\mathcal{S}_3(c_3)}^T ]^T $. By considering the joint decomposition of lower-dimensional discretized PDFs, we have constructed a single virtual non-negative CPD model and therefore uniqueness properties hold.  For example, by setting ${\mathcal{S}_1 = \{1,\dots, \lfloor\frac{N-1}{2}\rfloor -1 \}}$, ${\mathcal{S}_2 = \{ \lfloor \frac{N-1}{2} \rfloor,\dots,N-1 \}}$, $\mathcal{S}_3 = \{N \}$ we have that
\begin{align*}
&\underline{\mathbf{Y}}^{(1)}
 =
 \left(
\begin{bmatrix}
\mathbf{A}_{ \lfloor \frac{N-1}{2} \rfloor  } \\
\vdots \\
\mathbf{A}_{N-1}
\end{bmatrix} \odot
\begin{bmatrix}
\mathbf{A}_{1} \\
\vdots \\
\mathbf{A}_{ \lfloor \frac{N-1}{2} \rfloor -1}
\end{bmatrix}
\right) {\rm{diag}(\boldsymbol{\lambda})}\mathbf{A}_N^T.
\end{align*}
According to Theorem~\ref{thm:leurgans}, the CPD can be computed exactly if $R \leq (\lfloor \frac{N-1}{2} \rfloor -1) I$. Similarly, it is easy to verify that by setting $c_1 = c_2 = \lfloor  \frac{N}{3} \rfloor I$, i.e.,    $\alpha=\lfloor \log_2( \lfloor  \frac{N}{3} \rfloor I )\rfloor$, the CPD of   $\underline{\mathbf{Y}}$  is generically  unique for $R \leq 2 ^{2(\alpha-1)}$ according to Theorem~\ref{thm:generic}. The later inequality is implied by ${R \leq  \frac{(\lfloor{\frac{N}{3}}\rfloor I+1)^2}{16}}$ which shows that the bound is quadratic in $N$ and $I$.



\textbf{Remark 1}: The previous discussion suggests that finer discretization can lead to improved identifiability results. The number of hidden components may be arbitrarily large and we may still be able to identify the discretized conditional PDFs by increasing the dimensions  of the sub-tensors i.e., the discretization intervals of the random variables. The caveat is that one will need many more samples to reliably estimate these histograms. Ideally, one would like to have the minimum number of intervals that can guarantee identifiability of the conditional PDFs.

\textbf{Remark 2}: The factor matrices can be recovered by decomposing the lower-order probability tensors of dimension $N\geq 3$. It is important to note that histograms of subsets of two variables correspond to Non-negative Matrix Factorization (NMF) which is not identifiable unless additional conditions such as sparsity are assumed for the latent factors~\citep{fu2018}. Therefore, second-order distributions are not sufficient for recovering dense latent factor matrices.

\subsection{Recovery of the Conditional PDFs}
In the previous section we have shown that given lower-dimensional discretized PDFs, we can uniquely identify and recover discretized versions of the conditional PDFs via joint tensor decomposition. Recovering the true conditional PDFs from the discretized counterparts can be viewed as a signal reconstruction problem. We know that this is not possible unless the signals have some smoothness properties. We will use the following result.
\begin{Prop}
A PDF that is (approximately) band-limited with cutoff frequency $\omega_c$ can be recovered from uniform samples of the associated CDF taken $\frac{\pi}{\omega_c}$ apart.
\end{Prop}
\textit{Proof}: Assume that the PDF $f_X$ is band-limited with cutoff frequency $\omega_c$ i.e., its Fourier transform 
$\mathcal{F}(\omega) = 0$, $\forall \; |\omega| \geq \omega_c$. Let $F_X$ denote the CDF of $f_X$, $F_X(x) = \int_{-\infty}^{x} f_X(t) dt$.
We can express the integration as a convolution of the PDF with a unit step function, i.e., $F_X(x) = \int_{-\infty}^{\infty} f_X(t) u(x - t) d\tau$. The Fourier transform of a convolution is the point-wise product of Fourier transforms. Therefore, we can express the Fourier transform $\mathcal{G}(\omega)$ of the CDF as
\begin{equation}
\mathcal{G}(\omega) = \pi \delta(\omega) \mathcal{F}(0) +  \frac{\mathcal{F}(\omega)}{j\omega},
\label{eq:fourier}
\end{equation}
where $\delta(\cdot)$ is the Dirac delta. 
From Equation~\eqref{eq:fourier}, it is clear that the CDF obeys the same band-limit as the PDF . From Shannon's sampling theorem we have that
\begin{equation}
F_{X}(x) = \sum_{n=-\infty}^{ \infty} F_X(n T) \; {\rm sinc}\left( \frac{x - nT}{T} \right),
\end{equation}
where $T = \frac{\pi}{\omega_c}$. The PDF can then be determined by differentiation, which amounts to linear interpolation of the CDF samples using the derivative of the sinc kernel. Note that for exact reconstruction of $f_X$ an infinite number of data points are needed. In signal processing practice we always deal with finite support signals which are only approximately band-limited; the point is that the bandlimited assumption is accurate enough to afford high-quality signal reconstruction. In our present context, a good example is the Gaussian distribution: even though it is of infinite extent, it is not strictly bandlimited (as its Fourier transform is another Gaussian); but it is approximately bandlimited, and that is good enough for our purposes, as we will see shortly.
\begin{figure}[t]
\centering
\includegraphics[width=1\linewidth]{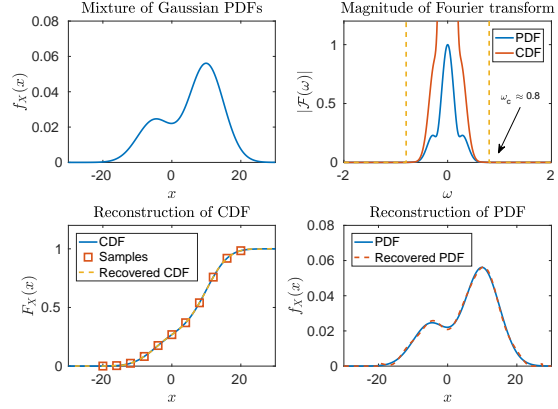}
\caption{Illustration of the key idea on a univariate Gaussian mixture. The CDF can be recovered from its samples if $T_s \leq \frac{\pi}{0.8}$.}
\label{fig:Gaussian}
\end{figure}
\begin{figure}[t]
\centering
\includegraphics[width= 0.8 \linewidth]{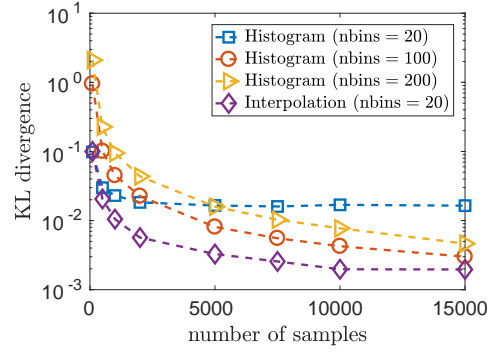}
\caption{KL divergence between the true mixture of Gaussians and different approximations.}
\label{fig:Gaussian_samples}
\end{figure}

In section~\ref{sec:ident}, we saw how lower-dimensional histograms can be used to obtain estimates of the discretized conditional PDFs. Now, consider the conditional PDF of the $n$-th variable given the $r$-th component. The corresponding column of factor matrix $\mathbf{A}_n$ is  
\begin{multline*}
\mathbf{A}_n[:,r] = [F_{X_n|H}(d_n^{1}|r) - F_{X_n|H}(d_n^{0}|r),\ldots, \\ 1 - F_{X_n|H}(d_n^{I-1}|r) ]^T.
\end{multline*}
Since, $F_{X_n|H}(d_n^{0}|r)=0$, we can compute
$F_{X_n|H}(d_n^{i}|r)$, $\forall i \in[I-1], n \in [N]$. We also know that $F_{X_n|H}(x_n|r)=1$, $ \forall x_n \geq d_n^I$. Therefore, we can recover the conditional CDFs using the interpolation formula
\begin{equation}
\!\!\!\!\!\! F_{X|H}(x_n|r) = \sum_{k=-L}^{L} F_{X_n|H}(kT|r) \; {\rm sinc}\left( \frac{x - kT}{T} \right),
\label{eq:cdf}
\end{equation}
where $T = d_n^i - d_n^{i-1}$ and $L$ a large integer. The conditional PDF $f_{X_n|H}$ can then be recovered via differentiation.
\subsection{Toy example}
An example to illustrate the idea is shown in Figure~\ref{fig:Gaussian}. Assume that the PDF of a random variable is a mixture of two Gaussian distributions with means $\mu_1=-6$, $\mu_2=10$ and standard deviations $\sigma_1 = \sigma_2 = 5$.  
It is clear from Figure~\ref{fig:Gaussian} that $\mathcal{F}(\omega) \approx 0$ for $|\omega| \geq \omega_c = 0.8$ and therefore the PDF is approximately band-limited. The CDF has the same band-limit, thus, it can be recovered from points being $T = \frac{\pi}{\omega_c} \approx 4$ apart. In this example we have used only $10$ discretization intervals as they suffice to capture $99\%$ of the data. We use the finite sum formula of Equation~\eqref{eq:cdf} to recover the CDF and then we recover the PDF by differentiating the CDF. The recovered PDF essentially coincides with the true PDF given a few exact estimates of the CDF as shown in Figure~\ref{fig:Gaussian}.

Figure~\ref{fig:Gaussian_samples} shows the approximation error for different methods  when we do not have exact points of the CDF but estimate them from randomly drawn samples. We know that a histogram converges to the true PDF as the number of samples grows and the bin width goes to $0$ at appropriate rate. However, when the conditional PDF is smooth, the interpolation procedure using a few discretization intervals leads to a lower approximation error compared to plain histogram estimates as illustrated in the figure. 
\section{Algorithm}
In this section we develop an algorithm for recovering the latent factors of the CPD model given the histogram estimates of lower-dimensional PDFs (Alg.~\ref{Alg:prop}). We define the following optimization problem
\begin{equation}
\begin{aligned}
\!\!\!\!\! \underset{{\{ \mathbf{A}_n\}_{n=1}^{N},\boldsymbol{\lambda} }}{\text{min.}} \;  &  \sum_{j=1}^N  \sum_{k>j}^N  \sum_{\ell>k}^N  {\rm{D}} \left( \widehat{\underline{\mathbf{X}}}_{jk\ell},[\![ \boldsymbol{\lambda},\mathbf{A}_{j},\mathbf{A}_k,\mathbf{A}_{\ell} ]\!]_R \right ) \\
\text{s.t. \quad} & \boldsymbol{\lambda}\geq \mathbf{0}, {\mathbf{1}}^T\boldsymbol{\lambda}=1 \\
&\mathbf{A}_n \geq \mathbf{0}, \; n=1\ldots N \\
& {\mathbf{1}}^T \mathbf{A}_n = \mathbf{1}^T, \; n=1\ldots N
\end{aligned}
\label{opt:coupled2}
\end{equation}
where $\rm{D}(\cdot,\cdot)$ is a suitable metric. The Frobenious norm and  Kullback-Leibler (KL) divergence are considered in this work. For probability tensors $\underline{\mathbf{X}},\underline{\mathbf{Y}}$ we define
\begin{equation*}
{\rm{D}}_{ \rm{KL} }(\underline{\mathbf{X}},\underline{\mathbf{Y}}) \triangleq \sum_{i_1,i_2,i_3} \underline{\mathbf{X}}[i_1,i_2,i_3] \log  \frac{\underline{\mathbf{X}}[i_1,i_2,i_3]}{\underline{\mathbf{Y}}[i_1,i_2,i_3]}
\end{equation*}
\begin{equation*}
{\rm{D}}_{ \rm{FRO}}(\underline{\mathbf{X}}, \underline{\mathbf{Y}}) \triangleq \sum_{i_1,i_2,i_3} \bigl( \underline{\mathbf{X}} [i_1,i_2,i_3] -\underline{\mathbf{Y}} [i_1,i_2,i_3] \bigr)^2.
\end{equation*}
Optimization problem~\eqref{opt:coupled2} is non-convex and NP-hard in its general form. Nevertheless, sensible approximation algorithms can be derived, based on well-appreciated nonconvex optimization tools. The idea is to cyclically update the variables while keeping all but one fixed. By fixing all other variables and optimizing with respect to $\mathbf{A}_{j}$ we have
\begin{equation}
\underset{{\mathbf{A}_j \in \mathcal{C}}}{\text{min.}} \;  \sum_{k \neq j} \sum_{\substack{l\neq j \\ l > k}} {\rm{D}} \left( \underline{\mathbf{X}}^{(1)}_{jk\ell},
(\mathbf{A}_{\ell} \odot \mathbf{A}_{k}) \textrm{diag}(\boldsymbol{\lambda}) \mathbf{A}_{j}^T \right ),
\label{opt:subproblem1}
\end{equation}
where  $\mathcal{C} = \{ \mathbf{A}  \mid \mathbf{A}\geq 0, \mathbf{1}^T\mathbf{A} = \mathbf{1}^T\}$.
Problem~\eqref{opt:subproblem1} is convex and can be solved efficiently using Exponentiated Gradient (EG)~\citep{kivinen1997} -- which is a special case of mirror descent~\citep{beck2003}. At each iteration $\tau$ of mirror descent we update $\mathbf{A}_{j}^{\tau}$ by solving\\
\begin{equation*}
\mathbf{A}_{j}^{\tau} = \argmin_{\mathbf{A}_{j} \in \mathcal{C}}\langle\ \nabla f \bigl(\mathbf{A}_{j}^{\tau-1} \bigr), \mathbf{A}_j \rangle + \frac{1}{\eta_{\tau}} B_{\Phi} \bigl(\mathbf{A}_{j},\mathbf{A}_{j}^{\tau-1} \bigr)
\end{equation*}
where $B_{\Phi}(\mathbf{A},\widehat{\mathbf{A}}) = \Phi (\mathbf{A}) -  \Phi(\widehat{\mathbf{A}}) -  \langle\ \mathbf{A}- \widehat{\mathbf{A}}, \nabla \Phi(\widehat{\mathbf{A}}) \rangle$ is a Bregman divergence. Setting $\Phi$ to be the negative entropy $\Phi(\mathbf{A}) = \sum_{i,j} \mathbf{A}(i,j) \log \mathbf{A}(i,j)$, the update becomes
\begin{equation}
\mathbf{A}_{j}^{\tau} = \mathbf{A}_{j}^{\tau-1} \circledast  \exp \left( - \eta_{\tau} \nabla f \left (\mathbf{A}_{j}^{\tau-1} \right) \right),
\end{equation}
where $\circledast$ is the Hadamard product, followed by column normalization $\mathbf{A}_{j}^{\tau}[:,r] = \frac{\mathbf{A}_{j}[:,r]}{ 1^T \mathbf{A}[:,r]}$.
The optimization problem with respect to $\boldsymbol{\lambda}$ is the following  
\begin{equation}
\!\!\! \underset{{ \boldsymbol{\lambda} \in \mathcal{C}}}{\text{min.}} \; \sum_{j,k,\ell}   {\rm{D}} \left( 
{\rm{vec}} (\underline{\mathbf{X}}_{jk\ell} ) , (\mathbf{A}_{\ell} \odot \mathbf{A}_{k} \odot 
\mathbf{A}_j ) \boldsymbol{\lambda} \right ).
\label{opt:subproblem2}
\end{equation}
The update rules for $\boldsymbol{\lambda}$ are similar
\begin{equation}
\boldsymbol{\lambda}^{\tau} = \boldsymbol{\lambda}^{\tau-1} \circledast  \exp \left( - \eta_{\tau} \nabla f \left( \boldsymbol{\lambda}^{\tau-1} \right) \right).
\end{equation}
The step $\eta_\tau$ can be computed by the Armijo rule~\citep{bertsekas1999}.

\begin{algorithm}[t]
\caption{Proposed Algorithm}
\label{Alg:prop}
\begin{algorithmic}[1]
\STATEx \hspace*{-.4cm}
 \textbf{Input}: A dataset $\mathbf{D}\in \mathbb{R}^{M \times N}$
\STATE Estimate $\underline{\mathbf{X}}_{jk\ell} \; \forall j,k,\ell \in [N],\; \ell>k>j$ from data.
\STATE Initialize $\{\mathbf{A}_n\}_{n=1}^N$ and $\boldsymbol{\lambda}$.
\REPEAT  
 \FORALL{$n \in [N]$} 
 \STATE{Solve opt. problem~\eqref{opt:subproblem1} via mirror descent.}
 \ENDFOR
 \STATE{Solve opt. problem~\eqref{opt:subproblem2} via mirror descent.} 
\UNTIL{convergence criterion is satisfied}
 \FORALL{$n \in [N]$} 
  \STATE{Recover $f_{X_n|H}$ by differentiation using Eq.~\eqref{eq:cdf}}
 \ENDFOR
\end{algorithmic}
\end{algorithm}

\section{Experiments}

\subsection{Synthetic Data}

In this section, we employ synthetic data simulations to showcase the effectiveness of the proposed algorithm. Experiments are conducted on synthetic datasets  $\{\mathbf{x}_m \}^M_{m=1}$ of varying sample sizes, generated  from  $R$ component distributions. We set the number of variables to $N = 10$, and vary the number of components $R \in \{5,10\}$. We run the algorithms using $5$ different random initializations and for each algorithm keep the model that yields the lowest cost. We evaluate the performance of the algorithms  by calculating the KL divergence between the true and learned model, which is approximated using Monte Carlo integration. Specifically, we generate $\{\mathbf{x}_{m'} \}_{m'=1}^{M'}$ test points, $M'=1000$ drawn from the mixture and approximate the KL divergence  between the true and learned model by \[{\rm{D}}_{ \rm{KL} } \left(f_{\mathcal{X}}, \widehat{f}_{\mathcal{X}} \right) \approx \frac{1}{M'} \sum_{m'=1}^{M'} \log {f_{\mathcal{X}}(\mathbf{x}_{m'})}/{\widehat{f}_{\mathcal{X}}(\mathbf{x}_{m'})}.\] We also compute the clustering accuracy on the test points as follows. Each data point $\mathbf{x}_{m'}$ is first assigned to the component yielding the highest posterior probability ${ \widehat{c}_m = \argmax_c f_{H|\mathcal{X}}( c | \mathbf{x}_m)}$. Due to the label permutation ambiguity, the obtained components are aligned with the true components using the Hungarian algorithm~\citep{kuhn1955}. The clustering accuracy is then calculated as the ratio of wrongly labeled data points over the total number of data points.
 For each scenario, we repeat $10$ Monte Carlo simulations and report the average results. We explore the following settings for the conditional PDFs: (1) Gaussian (2) GMM with two components (3) Gamma and (4) Laplace. The mixing weights are drawn from a Dirichlet distribution $\omega \sim {\rm{Dir}}(\alpha_1,\ldots,\alpha_r)$ with $\alpha_r = 10 \; \forall r$. We emphasize that our approach does not use any knowledge of the parametric form of the conditional PDFs; it only assumes smoothness. 
\begin{figure}[t]
\centering
\includegraphics[width=0.49 \linewidth]{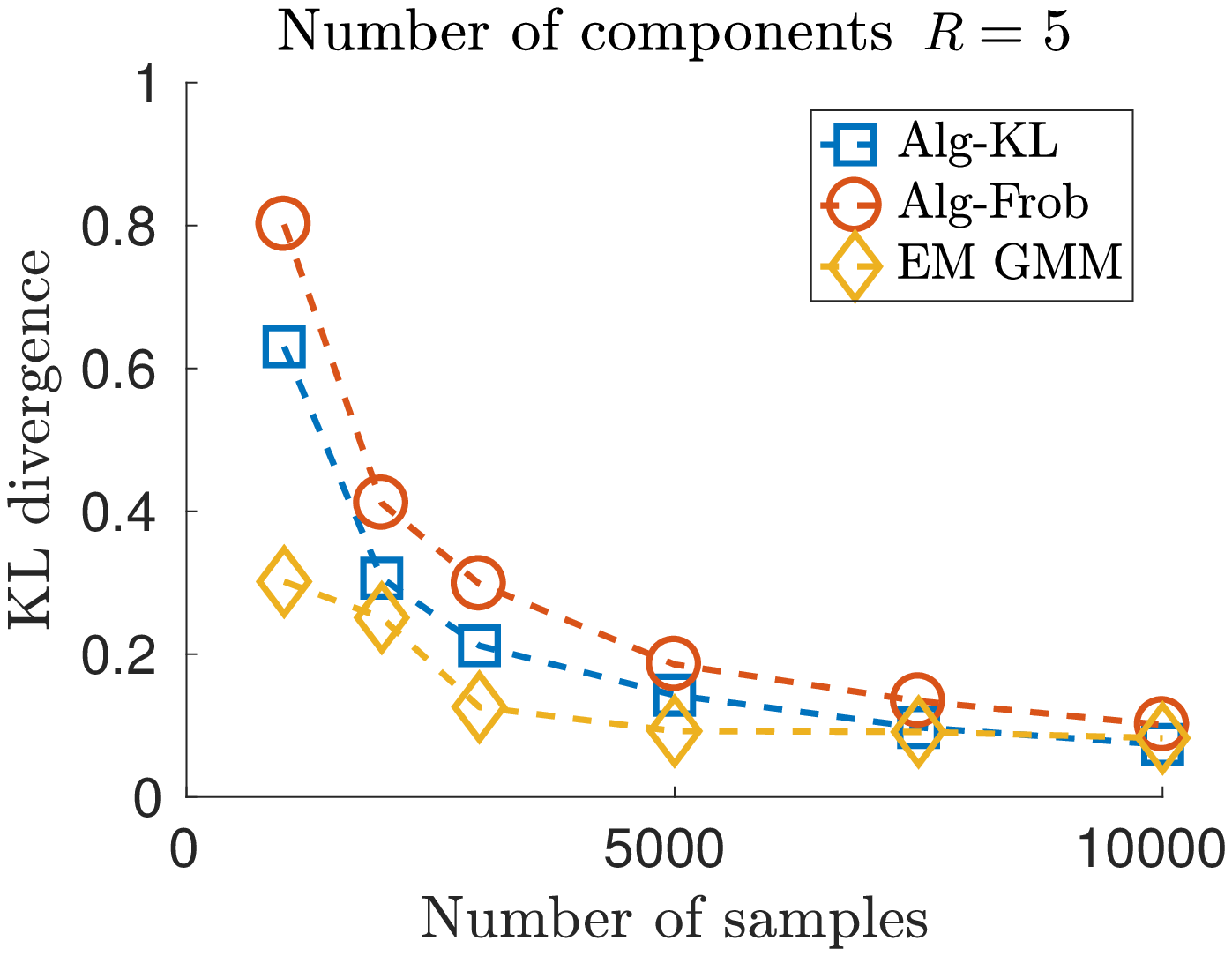}
\includegraphics[width=0.49 \linewidth]{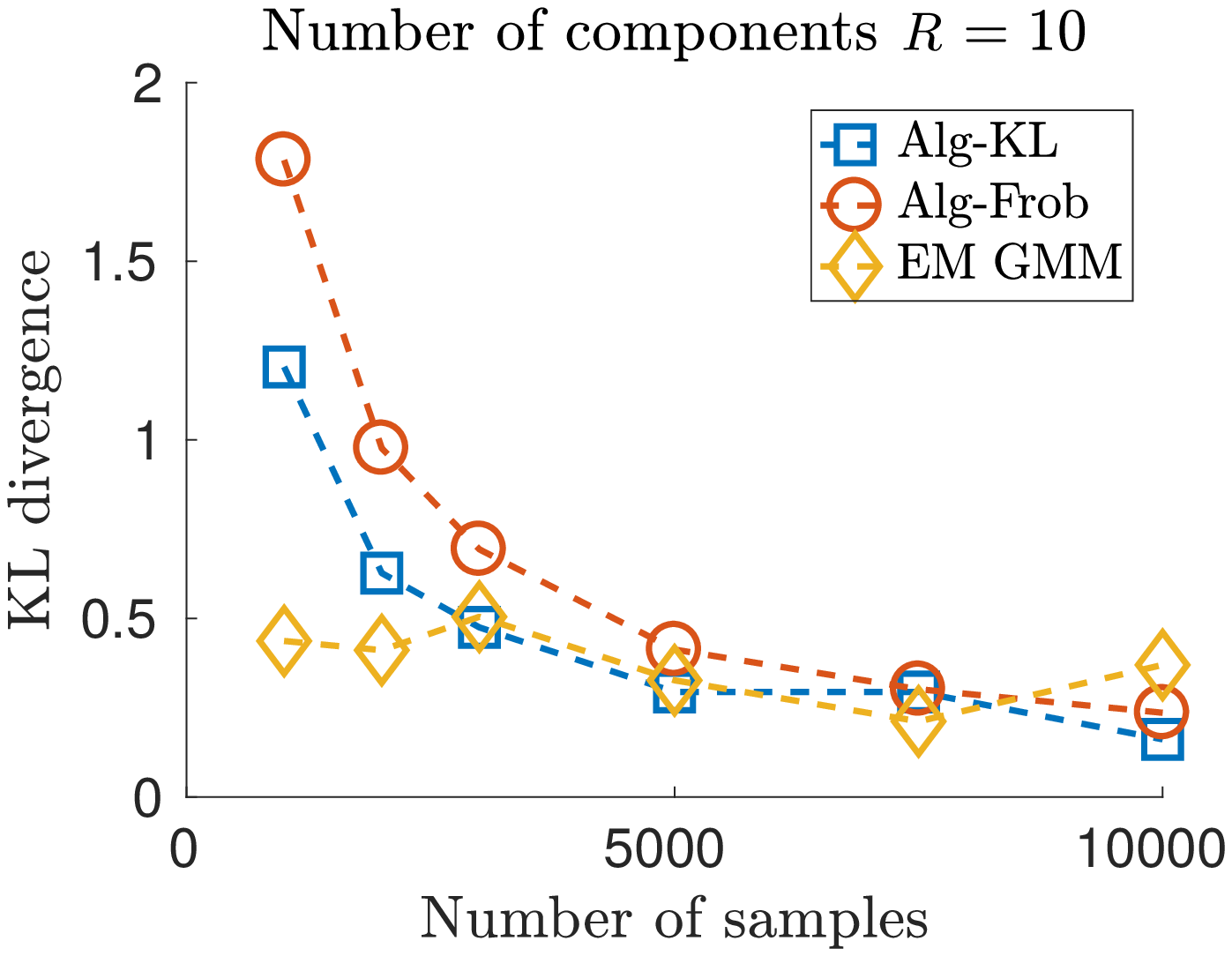}
\caption{KL divergence (Gaussian).}
\label{fig:kl_gaussian}
\end{figure}
\begin{figure}[t]
\centering
\includegraphics[width=0.49 \linewidth]{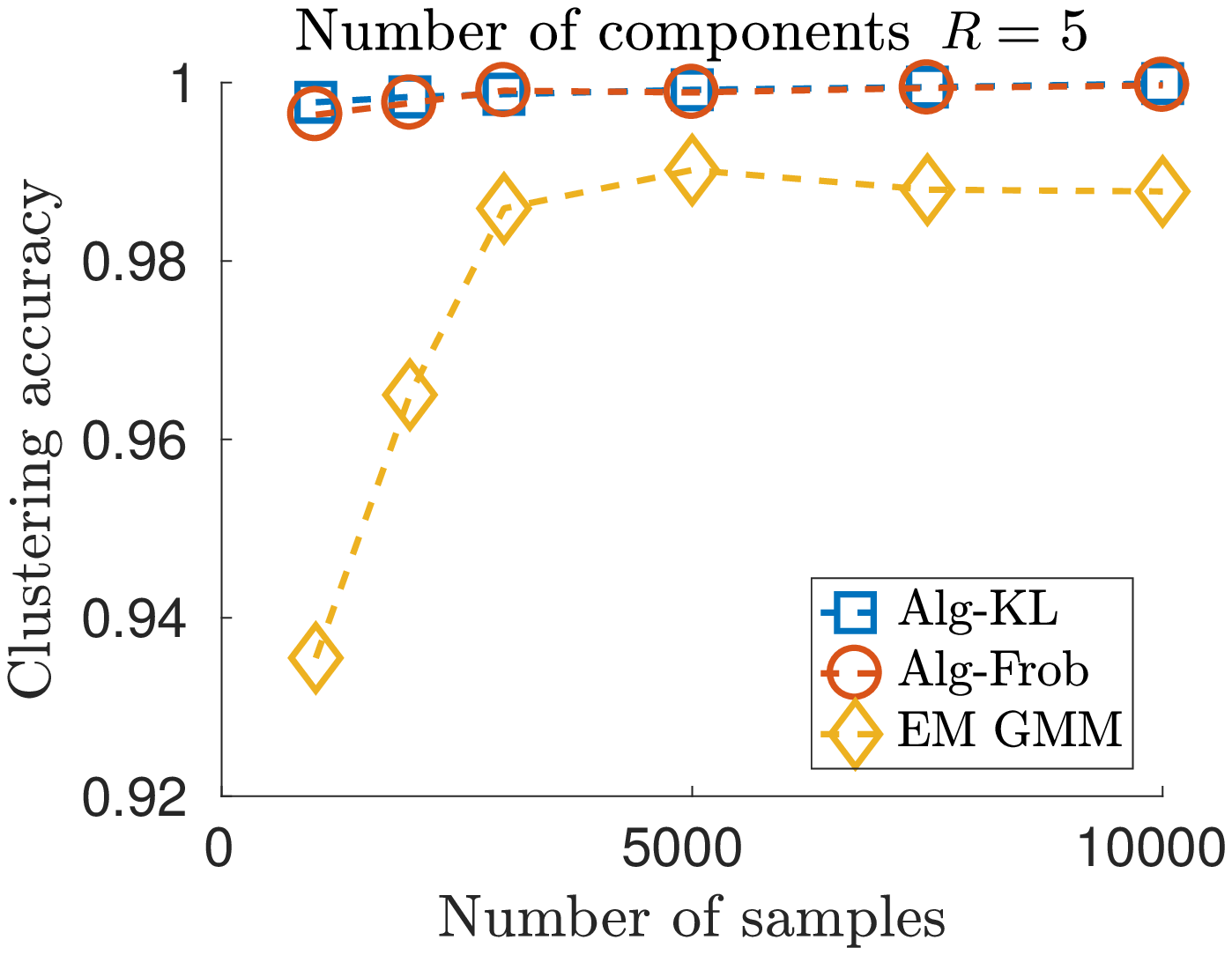}
\includegraphics[width=0.49 \linewidth]{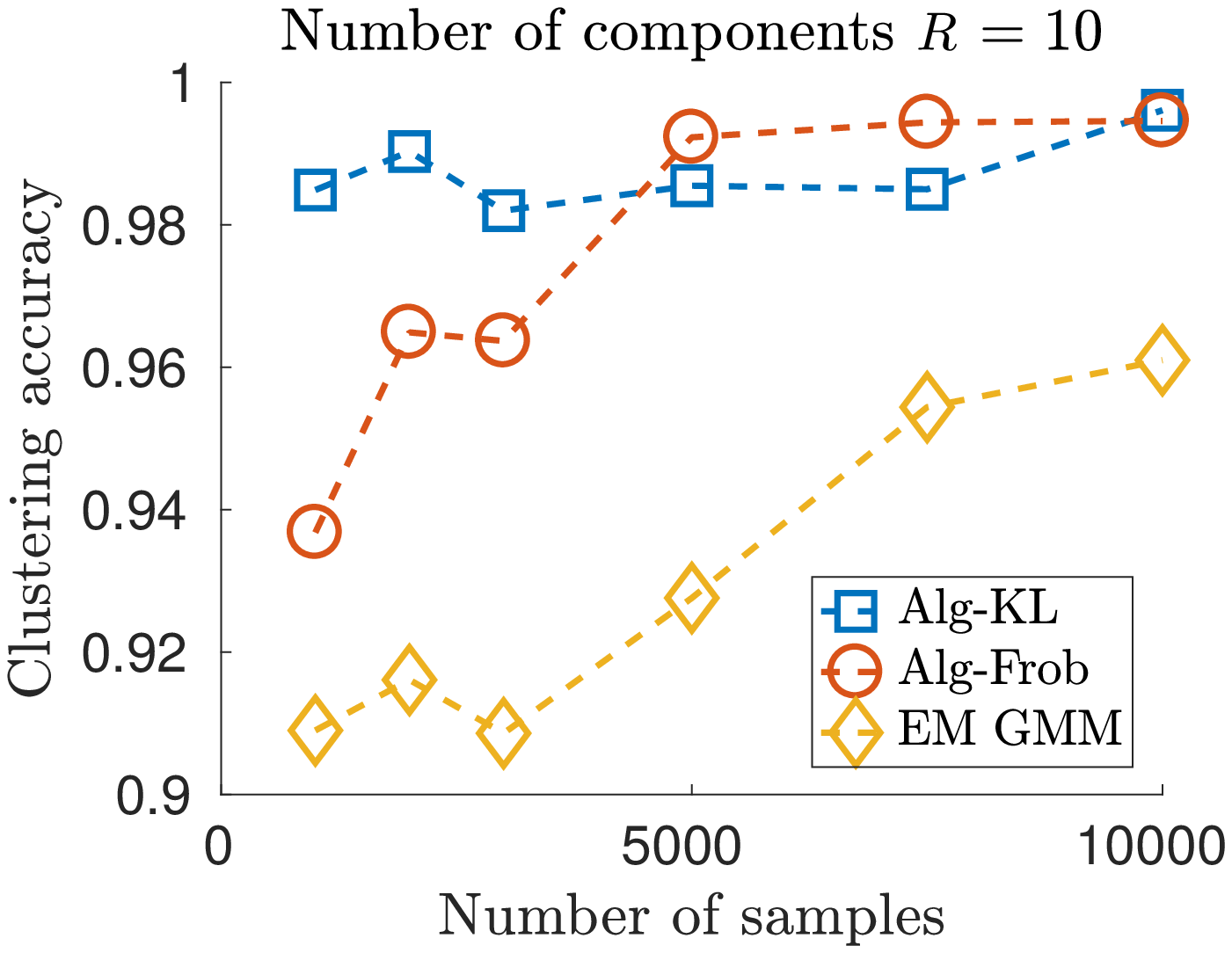}
\caption{Clustering accuracy (Gaussian).}
\label{fig:ac_gaussian}
\end{figure}

\begin{figure}[t]
\centering
\includegraphics[width=0.49 \linewidth]{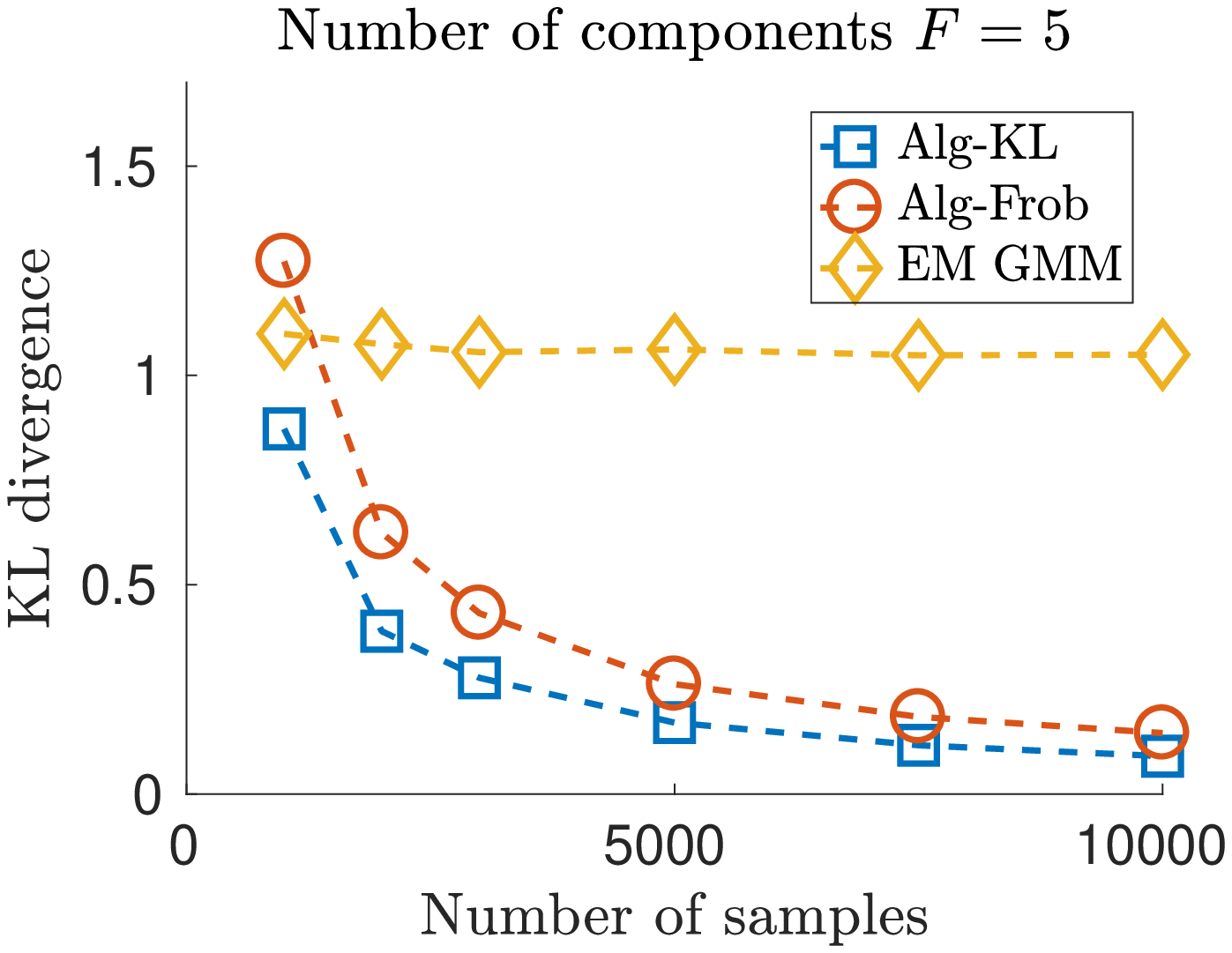}
\includegraphics[width=0.49 \linewidth]{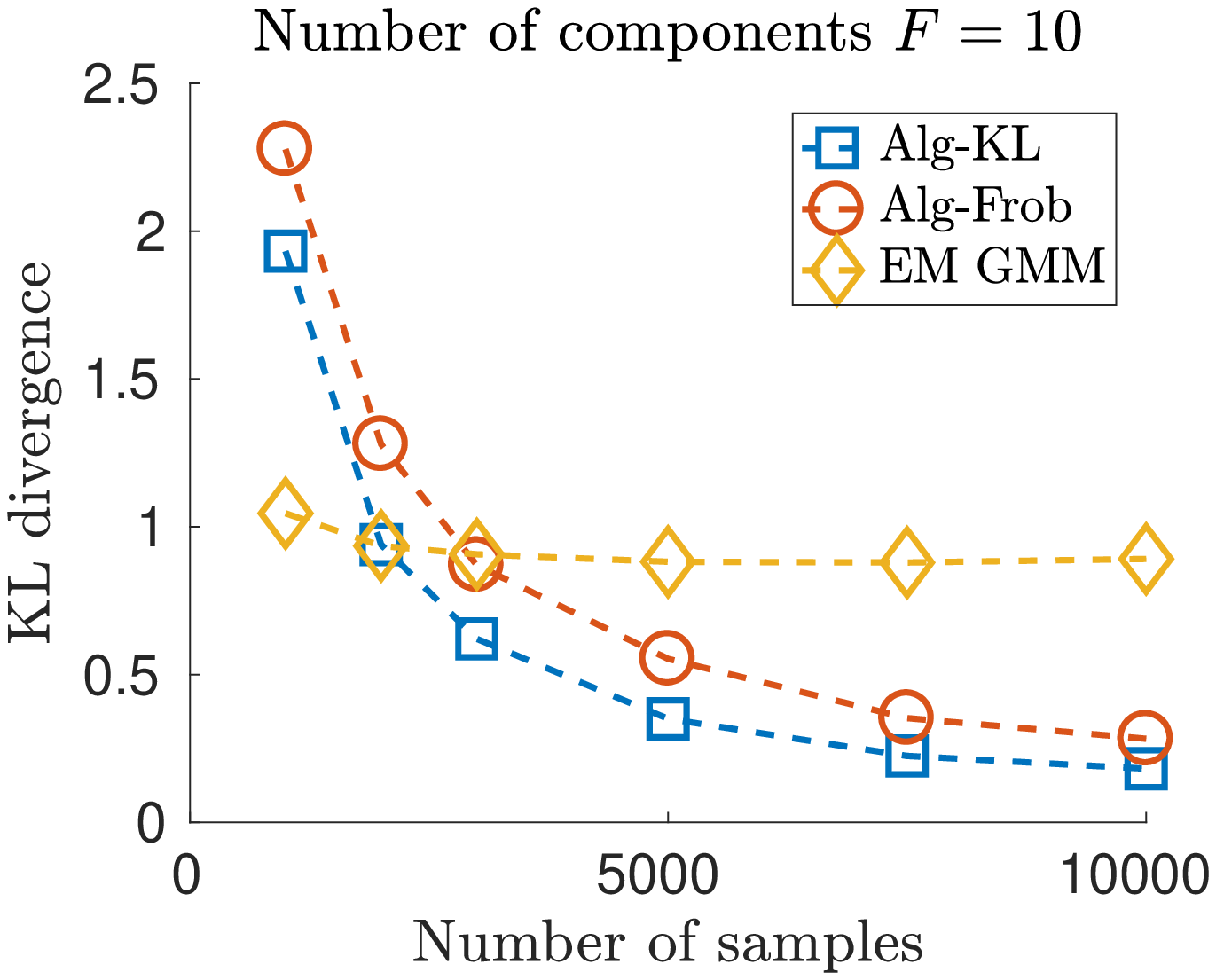}
\caption{KL divergence (GMM).}
\label{fig:kl_np}
\end{figure}

\begin{figure}[t]
\centering
\includegraphics[width=0.49 \linewidth]{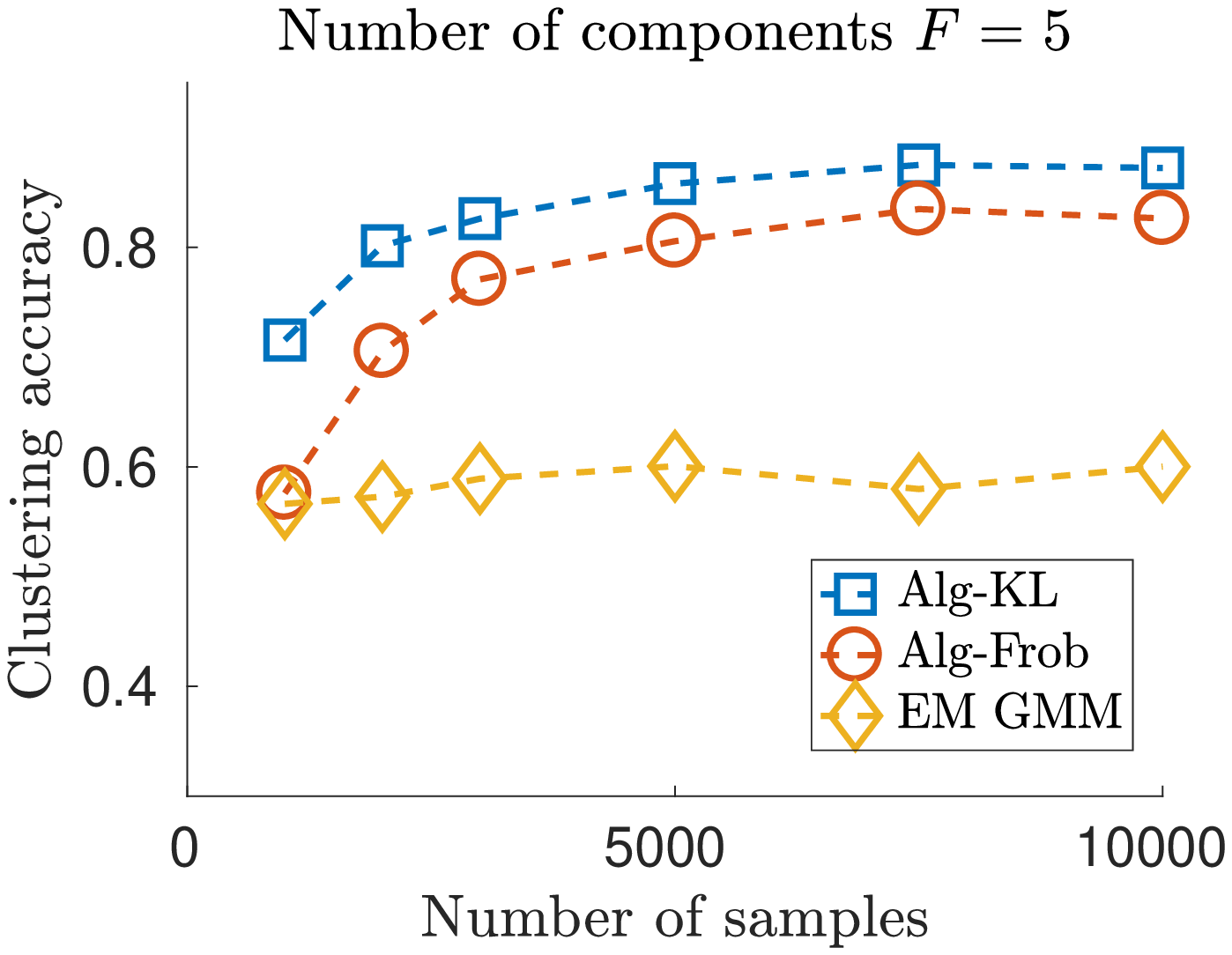}
\includegraphics[width=0.47 \linewidth]{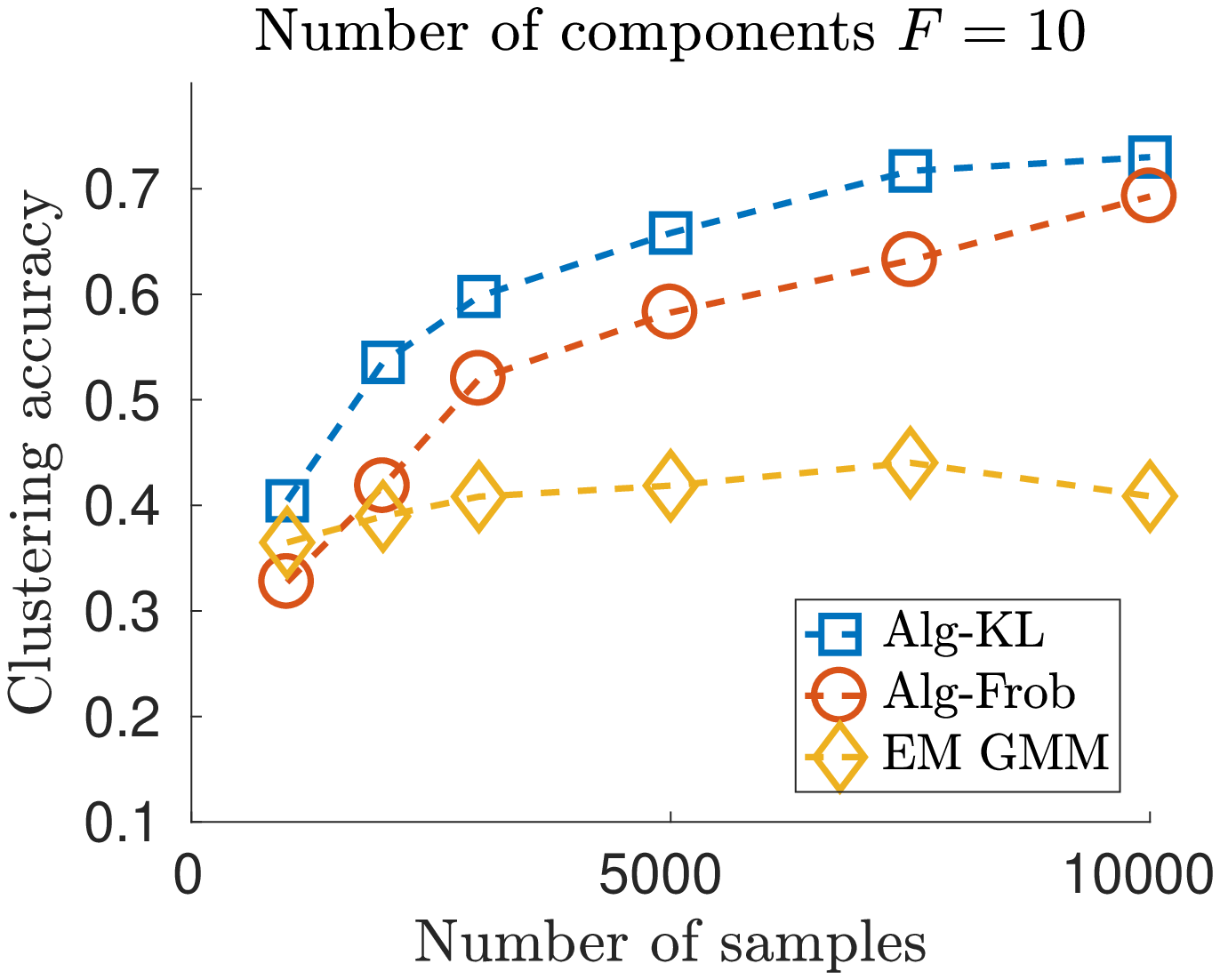}
\caption{Clustering Accuracy (GMM).}
\label{fig:ac_np}
\end{figure}

\textbf{Gaussian Conditional Densities}: In the first experiment we assume that each conditional PDF is a Gaussian. For cluster $r$ and random variable $X_n$ we set $f_{X_n|H}(x_n |r) = \mathcal{N}(\mu_{nr}, \sigma_{nr}^2)$. Mean and variance are drawn from uniform distributions, $\mu_{nr} \sim \mathcal{U}(-5,5) $,  $\sigma_{nr}^2 \sim \mathcal{U}(1,2)$. We compare the performance of our algorithms to that of EM (EM GMM). Figure~\ref{fig:kl_gaussian} shows the KL divergence between the true and the learned model for various dataset sizes and different number of components. We see that the  performance of our methods converges to that of EM despite the fact that we do not assume a particular model for the conditional densities. Interestingly, our approach performs better in terms of clustering accuracy as shown in Figure~\ref{fig:ac_gaussian}. We can see that although the joint distribution learned by EM is closer to the true in terms of the KL divergence, EM may fail to identify the true parameters of every component.

\begin{figure}[t]
\centering
\includegraphics[width=0.49 \linewidth]{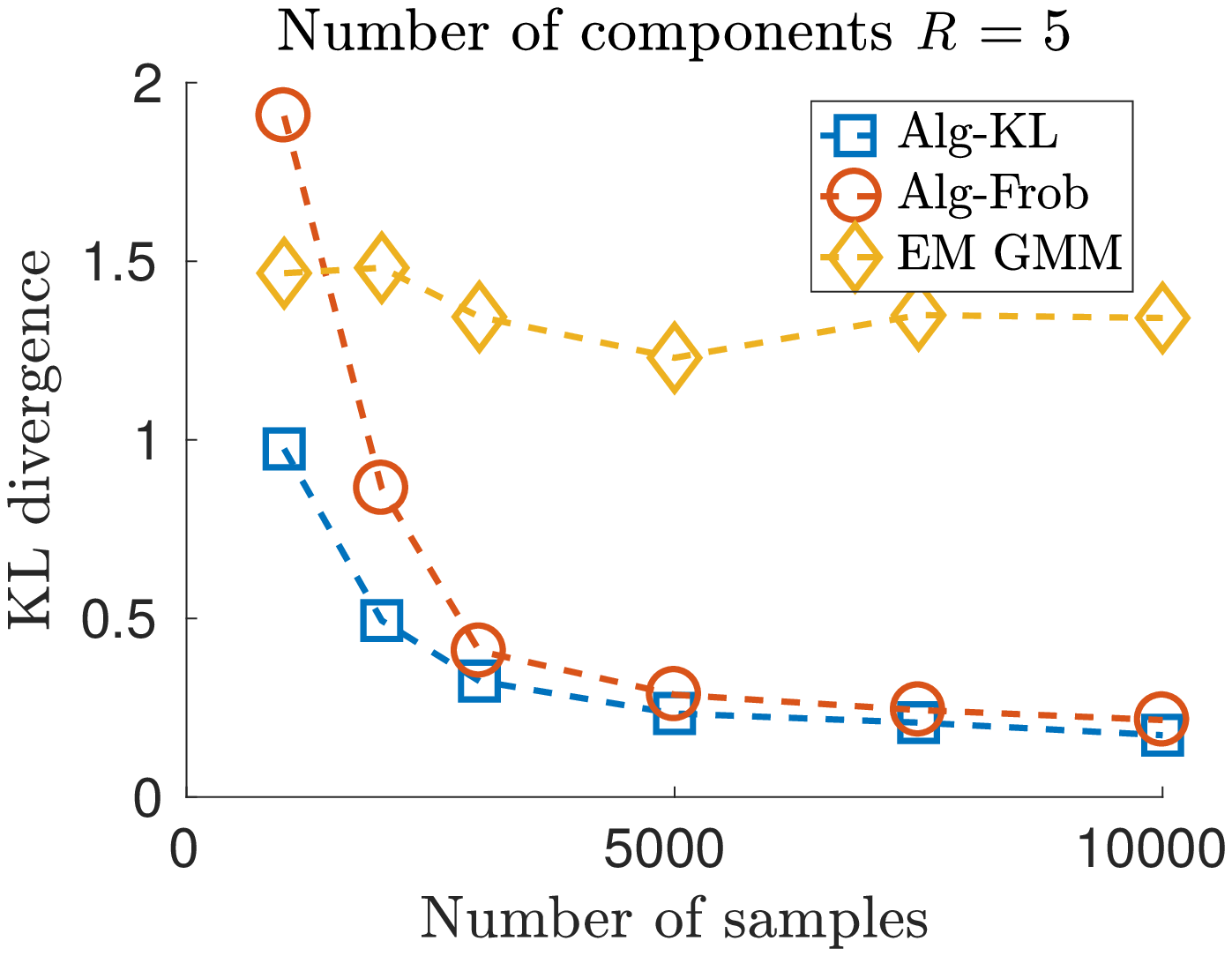}
\includegraphics[width=0.49 \linewidth]{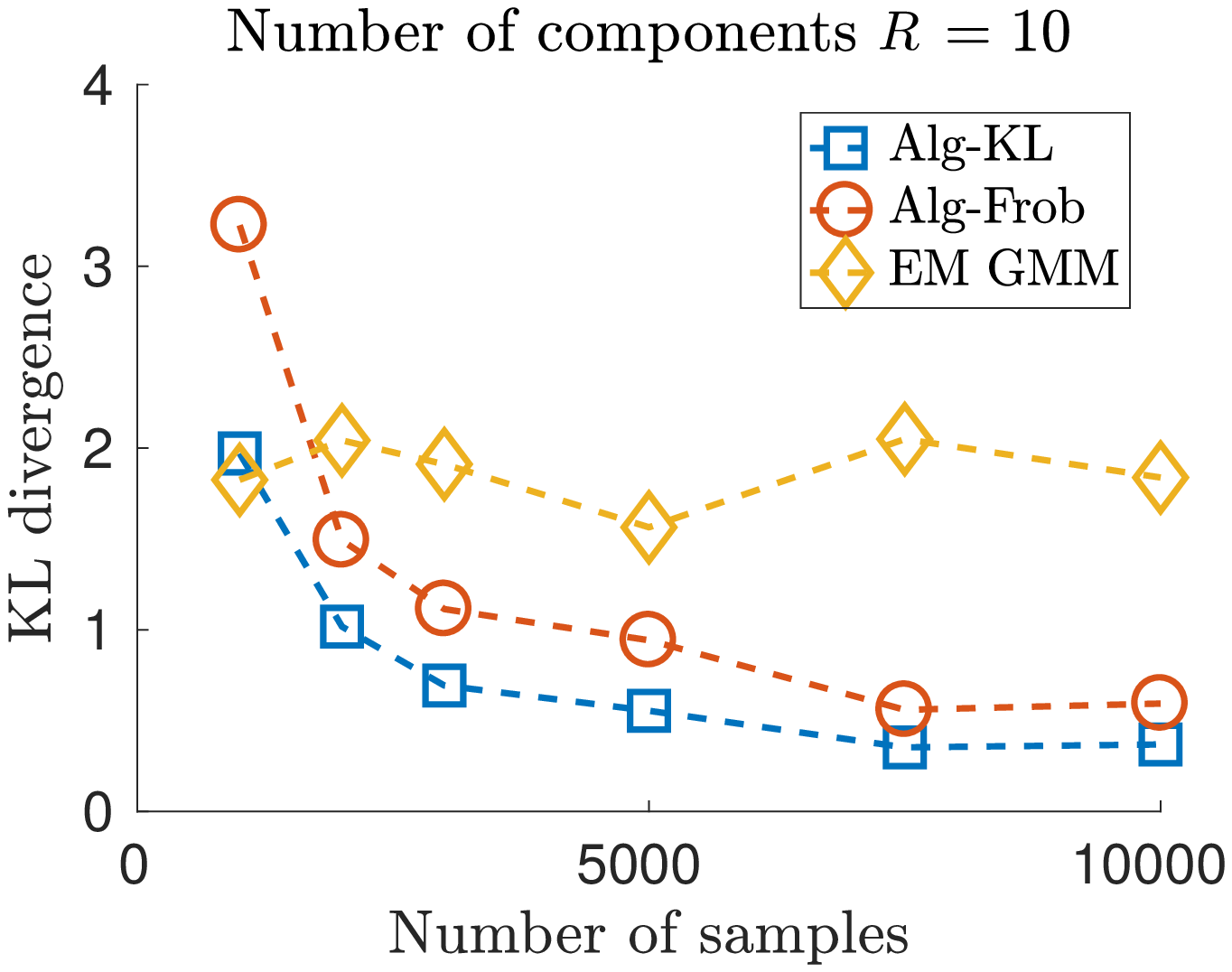}
\caption{KL divergence (Gamma).}
\label{fig:kl_gamma}
\end{figure}
\begin{figure}[t]
\centering
\includegraphics[width=0.49 \linewidth]{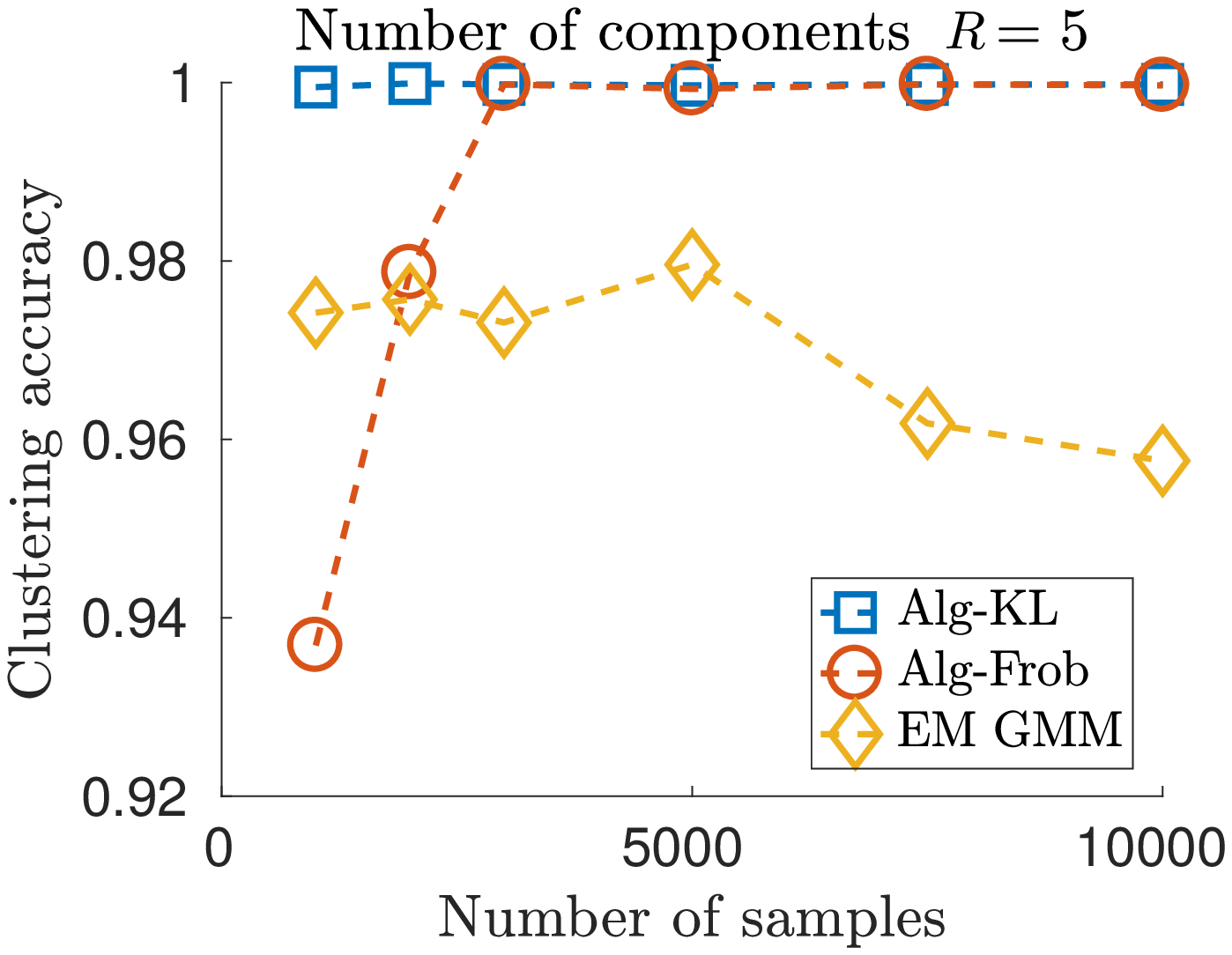}
\includegraphics[width=0.49 \linewidth]{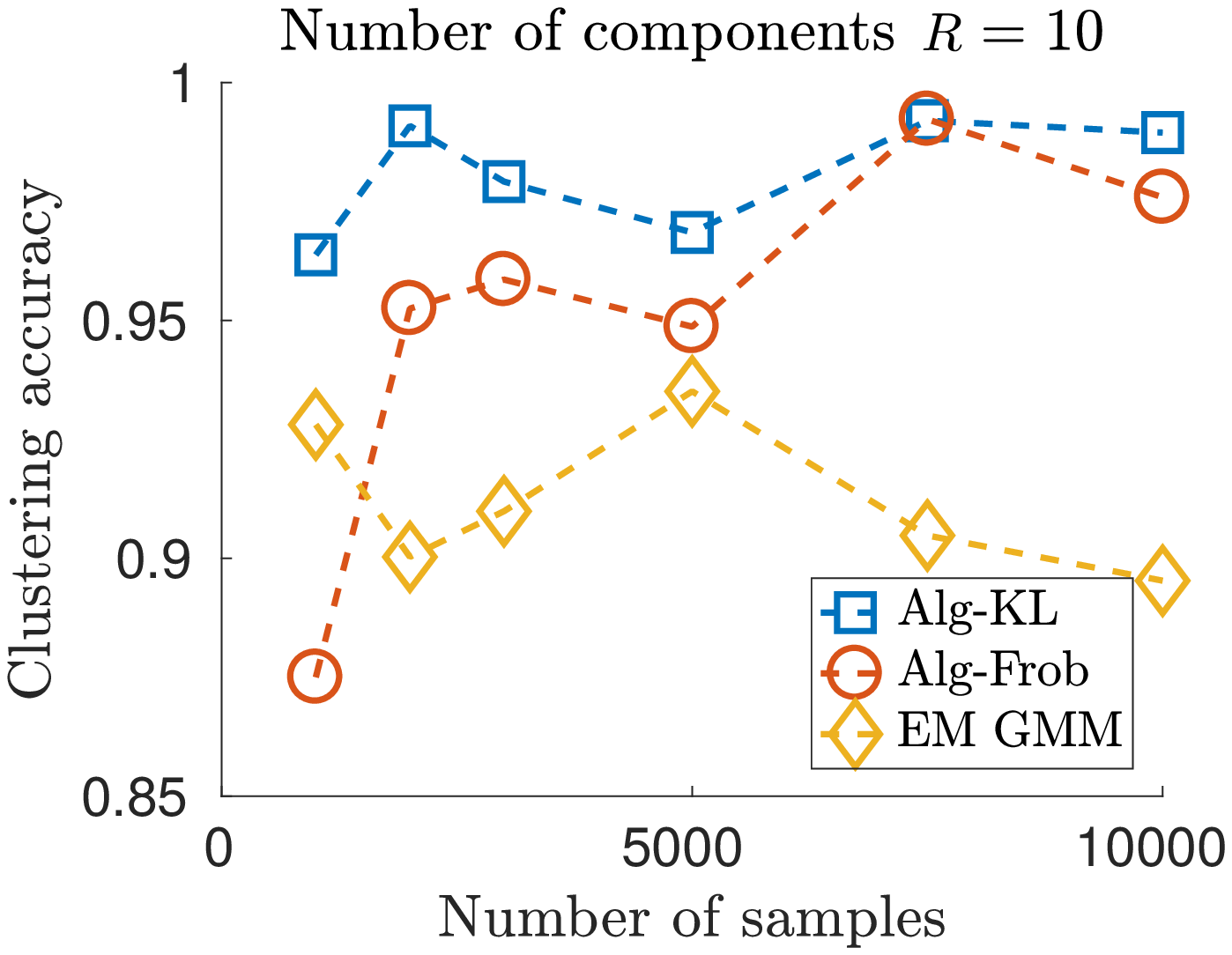}
\caption{Clustering accuracy (Gamma).}
\label{fig:ac_gamma}
\end{figure}

\begin{figure}[t]
\centering
\includegraphics[width=0.49 \linewidth]{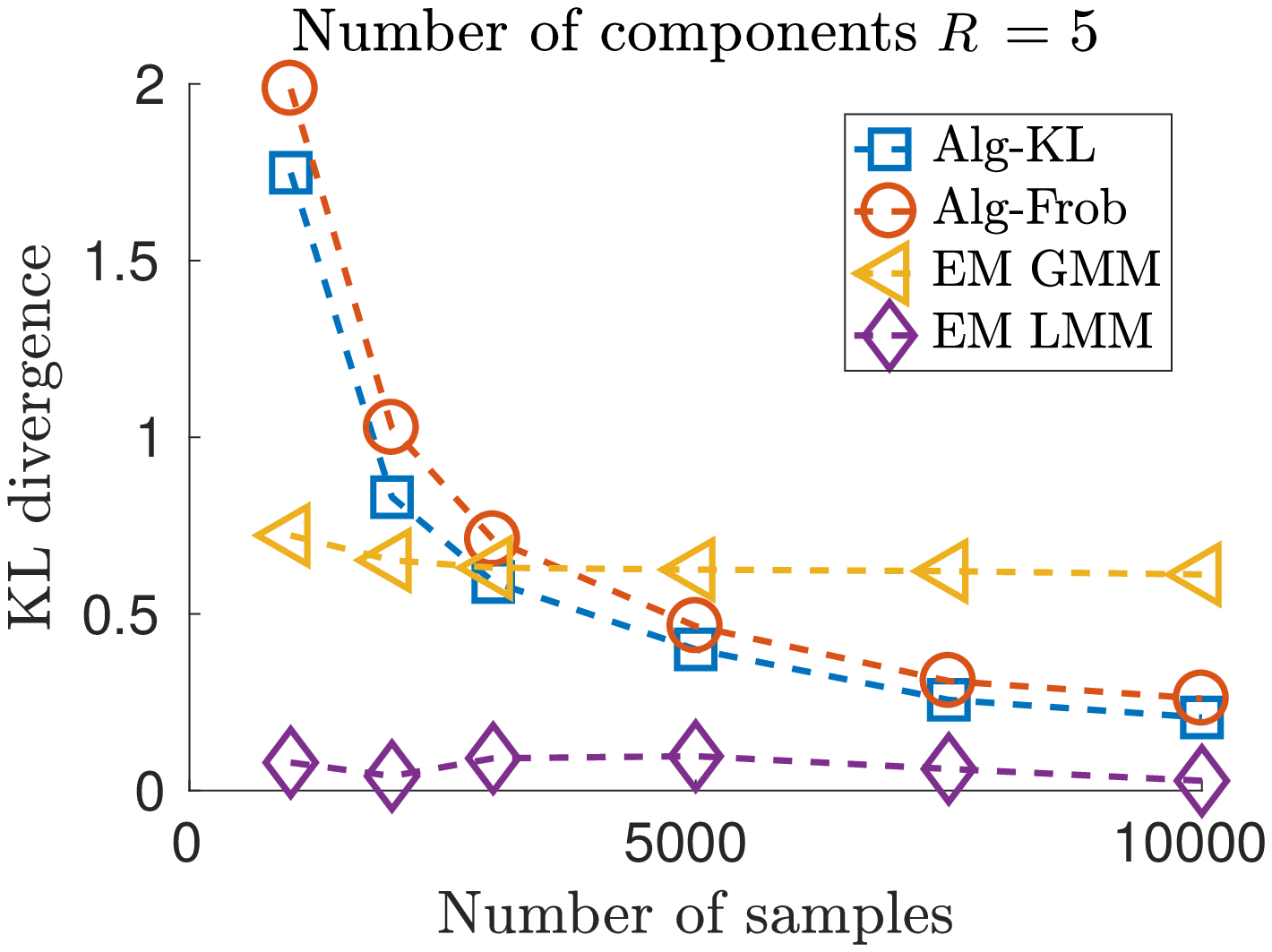}
\includegraphics[width=0.49 \linewidth]{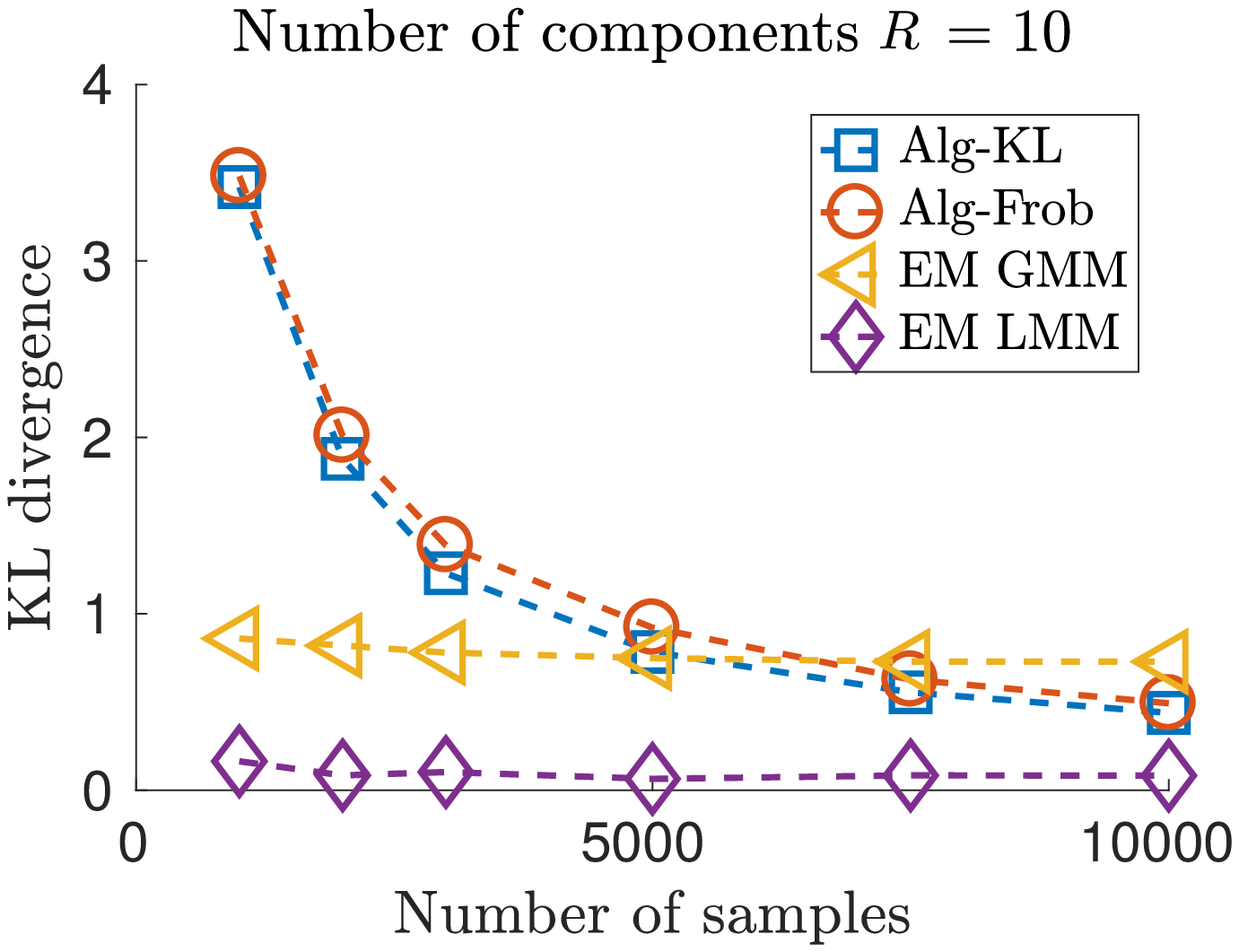}
\caption{KL divergence (Laplace).}
\label{fig:kl_laplace}
\end{figure}
\begin{figure}[t]
\centering
\includegraphics[width=0.49 \linewidth]{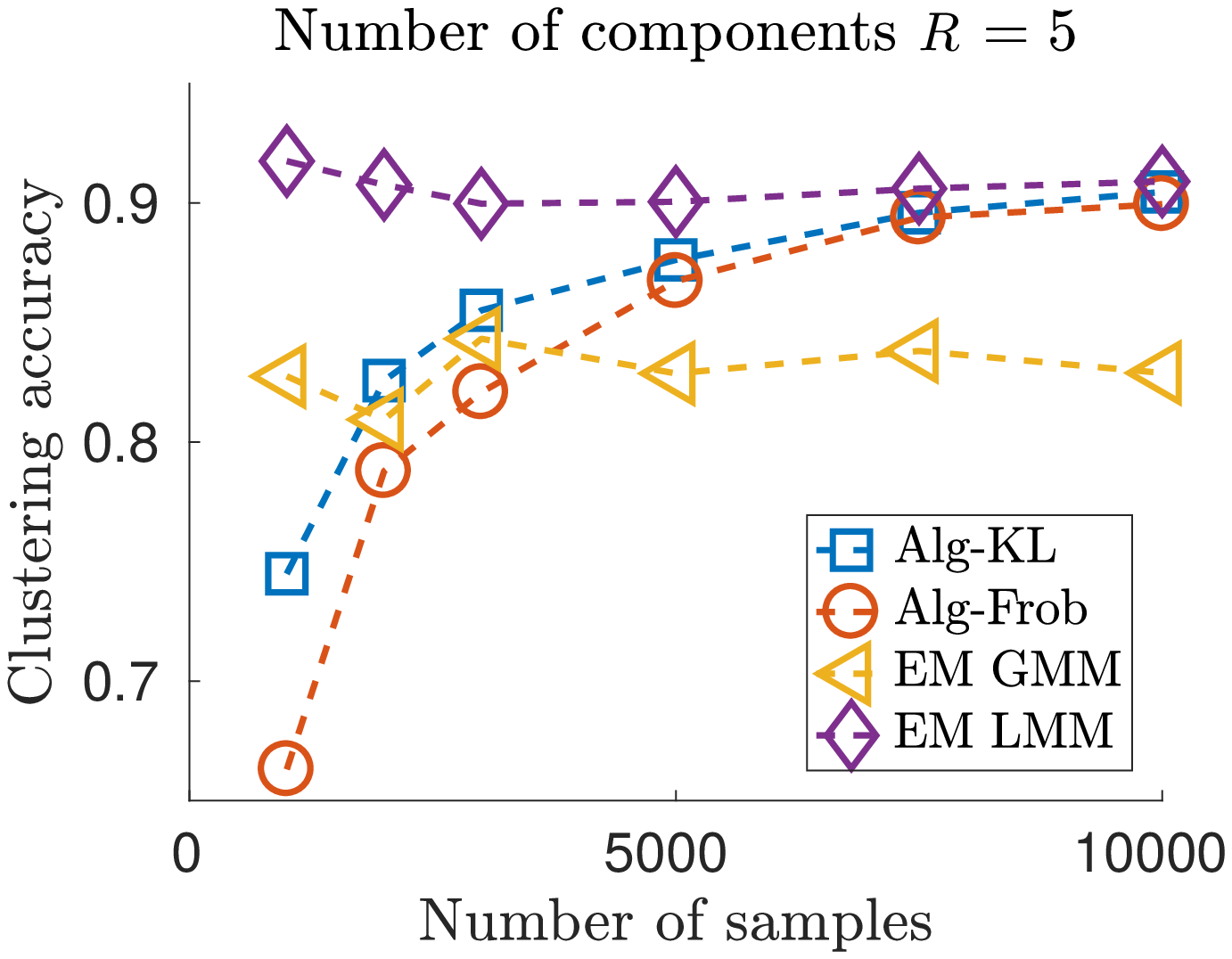}
\includegraphics[width=0.49 \linewidth]{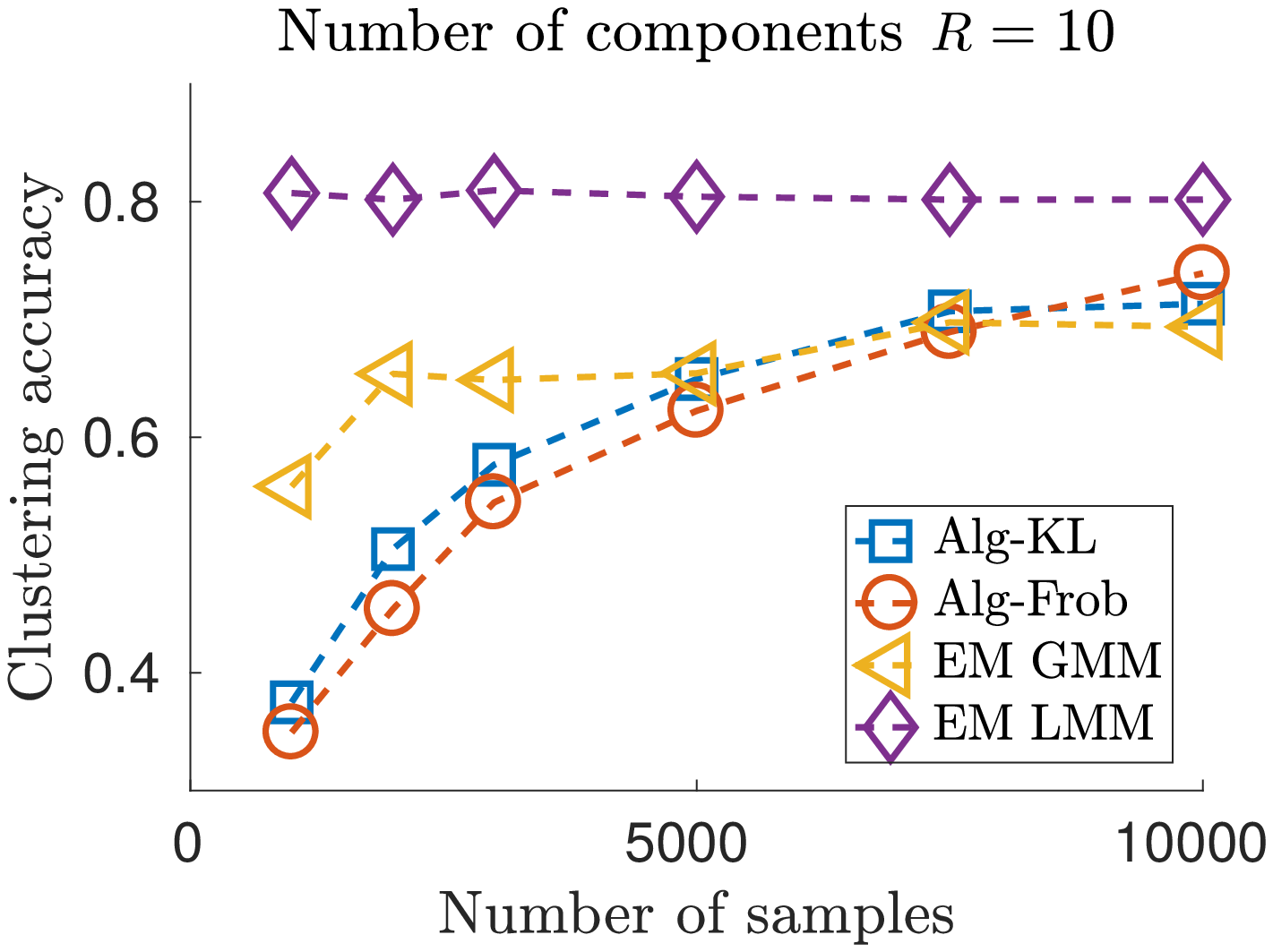}
\caption{Clustering accuracy (Laplace).}
\label{fig:ac_laplace}
\end{figure}

\textbf{GMM Conditional Densities}: In the second experiment we assume that each conditional PDF is itself a mixture model of two univariate Gaussian distributions. More specifically, we set ${f_{X_n|H}(x_n | r) = \frac{1}{2} \mathcal{N} \left (\mu^{(1)}_{nr}, {\sigma^{(1)2}_{nr}} \right) + \frac{1}{2} \mathcal{N} \left (\mu^{(2)}_{nr}, {\sigma^{(2)2}_{nr}} \right)}$.  Means and variances are drawn from uniform distributions $\mu_{nr}^{(1)} \sim \mathcal{U}(0,7) $,  $\sigma_{nr}^{(1)2} \sim \mathcal{U}(1,4)$, $\mu_{nr}^{(2)} \sim \mathcal{U}(-7,0) $,  $\sigma_{nr}^{(2)2} \sim \mathcal{U}(1,4)$. Our method is able to learn the mixture model in contrast to the EM GMM which exhibits poor performance, due to the model mismatch, as shown in Figures~\ref{fig:kl_np},~\ref{fig:ac_np}. 

\textbf{Gamma Conditional Densities}: Another example of a smooth distribution is the shifted Gamma distribution. We set $f_{X_n|H}(x_n| r) = \frac{1}{\beta^\alpha \Gamma(\alpha)  }  (x-\mu_{nr})^{\alpha-1} \exp( -\frac{x - \mu_{nr}}{\beta}) $ with $\alpha=5$, $\mu_{nr} \sim \mathcal{U}(-5,0) $,  $\beta_{nr} \sim \mathcal{U}(0.1,0.5)$. As the number of samples grows our method exhibits better performance, significantly outperforming EM GMM as shown in Figures~\ref{fig:kl_gamma},~\ref{fig:ac_gamma}.
\begin{figure}[h!]
\centering
\includegraphics[width=0.9 \linewidth]{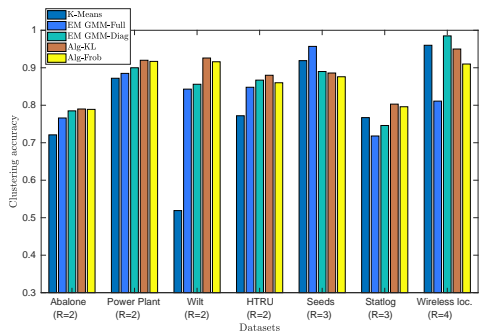}
\caption{Clustering accuracy on real datasets.}
\end{figure}

\textbf{Laplace Conditional Densities}: In the last simulated experiment we assume that each conditional PDF is a Laplace distribution with mean $\mu_{nr}$ and standard deviation $\sigma_{nr}$ i.e., $f_{X_n|H}(x_n | r) =  \frac{1}{\sqrt{2}\sigma_{nr}} \exp \left( \frac{\sqrt{2}| x_n - \mu_{nr}|}{\sigma_{nr}} \right)$. A Laplace distribution in contrast to the previous cases is not smooth (at its mean).  Parameters are drawn from uniform distributions, $\mu_{nf} \sim \mathcal{U}(-5,5) $,  $\sigma_{nf}^2 \sim \mathcal{U}(5,10)$. We compare the performance of our methods to that of the EM GMM and an EM algorithm for a Laplace mixture model (EM LMM). The proposed method approaches the performance of EM LMM and exhibits better performance in terms of KL and clustering accuracy compared to the EM GMM for higher number of data samples, as shown in Figures~\ref{fig:kl_laplace},~\ref{fig:ac_laplace}.

\subsection{Real Data}

Finally, we conduct several real-data experiments to test the ability of the algorithms to cluster data. We selected $7$ datasets with continuous variables suitable for classification or regression tasks from the UCI repository. For each labeled dataset we hide the label and treat it as the latent component. For datasets that contained a continuous variable as a response, we discretized the response into $R$ uniform intervals and treated it as the latent component. For each dataset we repeated $10$ Monte Carlo simulations by randomly splitting the dataset into three sets; $70 \%$  was used as a training set, $10\%$ as a validation set and $20\%$ as a test set. The validation set was used to select the number of discretization intervals which was either $5$ or $10$. We compare our methods against the EM GMM with diagonal covariance, EM GMM with full-covariance and the K-means algorithm in terms of clustering accuracy. Note that although the conditional independence assumption may not actually hold in practice, almost all the algorithms give satisfactory results in the tested datasets. The proposed algorithms perform well, outperforming the baselines in $5$ out of $7$ datasets while performing reasonably well in the remaining.

\section{Discussion and Conclusion}
We have  proposed a two-stage approach based on tensor decomposition and signal processing tools for recovering the conditional densities of mixtures of smooth product distributions. Our method does not assume a parametric form for the unknown conditional PDFs. We have formulated the problem as a coupled tensor factorization and proposed an alternating-optimization algorithm. Experiments on synthetic data have shown that when the underlying conditional PDFs are indeed smooth our method can recover them with high accuracy. Results on real data have shown satisfactory performance on data clustering tasks.
\bibliography{lib}
\end{document}